\documentclass[article,12pt,authoryear]{elsarticle}

\usepackage{lineno,hyperref}
\modulolinenumbers[5]

\usepackage{graphicx}
\usepackage{setspace}
\usepackage[T1]{fontenc}
\usepackage{textcomp}
\usepackage{times}
\usepackage{pifont}
\usepackage{listings} 
\usepackage[absolute]{textpos}
\usepackage{amsmath}
\usepackage{eurosym}
\usepackage{graphicx}
\usepackage{setspace}
\usepackage[T1]{fontenc}
\usepackage{textcomp}
\usepackage{times}
\usepackage{pifont}
\usepackage{listings} 
\usepackage[absolute]{textpos}
\usepackage{amsmath}
\usepackage{eurosym}
\usepackage{amsmath}
\usepackage{amsfonts}
\usepackage{nicefrac}
\usepackage{booktabs}
\usepackage{subcaption}
\usepackage{amsfonts,dcolumn}
\usepackage[table]{xcolor}
\usepackage{float}
\usepackage{cellspace}

\usepackage{amsfonts,dcolumn}
\usepackage{xcolor}
\usepackage{float}
\usepackage{xcolor,colortbl}
\usepackage{pgfplots}
\pgfplotsset{compat=1.18}
\usepackage{multirow}
\usepackage{arydshln}
\usepackage{enumitem}

\usepackage[toc, automake]{glossaries} %
\setkeys{glslink}{hyper=false}
\newacronym{AdaGrad}{AdaGrad}{Adaptive Gradient algorithm}
\newacronym{AI}{AI}{Artificial Intelligence}
\newacronym[longplural={Auxiliary Classification Generative Adversarial Networks}]{AC-GAN}{AC-GAN}{Auxiliary Classification Generative Adversarial Network}
\newacronym{Adam}{Adam}{Adaptive Moment Estimation}
\newacronym{API}{API}{Application Programming Interface}
\newacronym{ARIMA}{ARIMA}{Auto-Regressive Integrated Moving Average}
\newacronym[longplural={Autoencoders}]{AE}{AE}{Autoencoder}
\newacronym[longplural={Artificial Neural Networks}]{ANN}{ANN}{Artificial Neural Network}
\newacronym{AFE}{AFE}{Automated Feature Engineering}
\newacronym{BPTT}{BPTT}{backpropagation through time}
\newacronym{ConvLSTM2D}{ConvLSTM2D}{Bi-dimensional Convolutional LSTM}
\newacronym[longplural={Convolutional Neural Networks}]{CNN}{CNN}{Convolutional Neural Network}
\newacronym{DL}{DL}{Deep Learning}
\newacronym[longplural={Deep Neural Networks}]{DNN}{DNN}{Deep Neural Network}
\newacronym{DFT}{DFT}{Discrete Fourier Transform}
\newacronym{DRL}{DRL}{Deep Reinforcement Learning}
\newacronym{XOR}{XOR}{Exclusive Or}
\newacronym{FE}{FE}{Feature Engineering}
\newacronym{FFT}{FFT}{Fast Fourier Transform}
\newacronym[longplural={Feed Forward Neural Networks}]{FFNN}{FFNN}{Feed Forward Neural Network}
\newacronym{FIFO}{FIFO}{First In First Out}
\newacronym[longplural={Generative Adversarial Networks}]{GAN}{GAN}{Generative Adversarial Network}
\newacronym[longplural={Genetic Algorithms}]{GA}{GA}{Genetic Algorithm}
\newacronym{HAR}{HAR}{Human Activity Recognition}
\newacronym{LVQ}{LVQ}{Learning Vector Quantization}
\newacronym{LSTM}{LSTM}{Long Short-Term Memory}
\newacronym{ML}{ML}{Machine Learning}
\newacronym{MSE}{MSE}{Mean Square Error}
\newacronym{MTS}{MTS}{Multivariate Time Series}
\newacronym[longplural={Neural Networks}]{NN}{NN}{Neural Network}
\newacronym{OVA}{OVA}{One-vs-All}
\newacronym{OVO}{OVO}{One-vs-One}
\newacronym{PSO}{PSO}{Particle Swarm Optimization}
\newacronym{PM}{PM\textsubscript{2.5}}{Particulate Matter 2.5}
\newacronym{PCA}{PCA}{Principal Component Analysis}
\newacronym{PL}{PL}{Profit \& Loss}
\newacronym[longplural={Raw Data Frames}]{RDF}{RDF}{Raw Data Frame}
\newacronym{RTD}{RTD}{Raw Time Distributed}
\newacronym{ReLU}{ReLU}{Rectified Linear Unit}
\newacronym{RMSE}{RMSE}{Root Mean Square Error}
\newacronym{RMSprop}{RMSprop}{Root Mean Square Propagation}
\newacronym[longplural={Recurrent Neural Networks}]{RNN}{RNN}{Recurrent Neural Network}
\newacronym[longplural={Simple Recurrent NNs}]{SRN}{SRN}{Simple Recurrent NN}
\newacronym[longplural={Stacked Autoencoders}]{SAE}{SAE}{Stacked Autoencoder}
\newacronym{SGD}{SGD}{Stochastic Gradient Descent}
\newacronym[longplural={Support Vector Machines}]{SVM}{SVM}{Support Vector Machine}
\newacronym{TCN}{TCN}{Temporal Convolutional Network}
\newacronym{UCI}{UCI}{University of California Irvine}
\newacronym[longplural={Variational Autoencoders}]{VAE}{VAE}{Variational Autoencoder}
\newacronym{WoM}{WoM}{Weight Of Money}
\newacronym{WPI}{WPI}{Wholesale Price Index}

\makeglossaries

\usepackage{pdflscape}
\usepackage{rotating}
\usepackage{amssymb}
\usepackage[a4paper,portrait,includefoot,margin=0.8in]{geometry}
\usepackage{graphicx}
\usepackage{pdfpages}
\usepackage{endnotes}
\usepackage{rotating}%
\usepackage{setspace}
\usepackage{caption}
\usepackage{subcaption}
\usepackage{float}
\usepackage{pdflscape}[2016/05/14]
\usepackage{afterpage}
\usepackage{capt-of}
\usepackage{lipsum}
\usepackage{xcolor}
\usepackage{adjustbox}
\usepackage{eurosym}
\usepackage[most]{tcolorbox}%
\usepackage{amsmath}
\usepackage{amsthm}
\usepackage{mathtools}
\usepackage{amssymb}
\usepackage{float}
\usepackage{moreverb}
\usepackage{lineno}
\usepackage{subcaption}
\usepackage{caption}
\usepackage{setspace}
\usepackage{fancyhdr,lipsum}
\usepackage{graphicx}

\usepackage{natbib}
\usepackage{adjustbox}
\usepackage{nameref}
\usepackage{xurl}
\usepackage{hyperref}

\definecolor{mygreen}{RGB}{200, 220, 200}
\definecolor{myblue}{RGB}{184, 193, 218}
\definecolor{mylightblue}{RGB}{214, 220, 235}
\definecolor{mypink}{RGB}{221, 173, 202}
\definecolor{mylightpink}{RGB}{234, 209, 227}

\makeatletter
\DeclareRobustCommand{\rvdots}{%
	\vbox{
		\baselineskip2\p@\lineskiplimit\z@
		\kern-\p@
		\hbox{.}\hbox{.}\hbox{.}
}}
\makeatother

\biboptions{authoryear}

\begin{document}
\include{Xlatin1}
\begin{frontmatter}

\title{Convolutional Attention in Betting Exchange Markets}
\tnotetext[]{This work was financially supported by UID/00147 – Systems and Technologies Center (SYSTEC), with the support of the Associate Laboratory Advanced Production and Intelligent Systems (ARISE), LA/P/0112/2020 (DOI: 10.54499/LA/P/0112/2020), both funded by national funds through FCT/MCTES (PIDDAC).}

\author[rvt]{Rui Gon\c{c}alves}
\ead{rjpg@fe.up.pt}
\author[ave]{Vitor Miguel Ribeiro}
\ead{vsribeiro@fep.up.pt}
\author[rvt]{Roman Chertovskih}
\ead{roman@fe.up.pt}
\author[rvt]{Ant\'onio Pedro Aguiar}
\ead{apra@fe.up.pt}

\address[rvt]{
SYSTEC-ARISE Research Center for Systems and Technologies, Faculty of Engineering, University of Porto}
\address[ave]{Department of Economics, School of Economics and Management, Porto University of Porto}

\begin{abstract}
This study presents the implementation of a short-term forecasting system for price movements in exchange markets, using market depth data and a systematic procedure to enable a fully automated trading system. The case study focuses on the UK to Win Horse Racing market during the pre-live stage on the world's leading betting exchange, Betfair. Innovative convolutional attention mechanisms are introduced and applied to multiple recurrent neural networks and bi-dimensional convolutional recurrent neural network layers. Additionally, a novel padding method for convolutional layers is proposed, specifically designed for multivariate time series processing. These innovations are thoroughly detailed, along with their execution process. The proposed architectures follow a standard supervised learning approach, involving model training and subsequent testing on new data, which requires extensive pre-processing and data analysis. The study also presents a complete end-to-end framework for automated feature engineering and market interactions using the developed models in production. The key finding of this research is that all proposed innovations positively impact the performance metrics of the classification task under examination, thereby advancing the current state-of-the-art in convolutional attention mechanisms and padding methods applied to multivariate time series problems. 
\end{abstract}


\end{frontmatter}


\section{Introduction}

Automated trading systems have revolutionized financial markets, with applications that span various domains, including exchange markets. One particularly interesting and evolving area is the betting exchange market, where predicting price movements can lead to profitable strategies. This study focuses on implementing a fully integrated short-term forecasting system to predict future price movements in exchange markets. To meet this purpose, \gls{AI} automated trading agents are developed to execute trades in the United Kingdom (UK) to Win Horse Racing market during the pre-live stage on Betfair, the world's largest betting exchange platform. These agents operate within a 10-minute window before a race begins, where odds (i.e., prices) are subject to speculation, having in mind the objective to generate profit by buying and selling bets at different odds.

The focus of this study is on the 10-minute window before a race begins, with the goal of accurately classifying the price movement in the final 2 minutes before the race begins, based on information from previous minutes. There is no sliding window overlap, meaning each race serves as an individual example. By combining prediction models with trading strategies, this study evaluates the performance of the entire system and assesses the combined impact of individual components, as suggested by \cite{RGoncalves:2013} and \cite{GONCALVES201938}. A complete end-to-end framework has been developed, involving the preparation of complex market data, predictive model training, and the deployment of several \gls{DL} \gls{NN} models within a fully automated trading system. The software implementation framework collects market depth data, performs \gls{FE}, and operationalizes \gls{DL} \gls{NN} architectures for real-time use in trading simulations. Currently, the existing literature lacks end-to-end frameworks, with most research focusing on basic learning processes for prediction and/or classification tasks. In these studies, authors often claim the superiority of their models, yet fail to fully integrate them within a complete operational framework. Therefore, this study extends beyond indicator analysis, model building, and the typical train/test learning process to include a comprehensive evaluation of the overall performance of the trading system.

Moreover, even for the specific task of training and testing \gls{DL} \gls{NN} models, the aim of this study is to advance the understanding of whether novel convolutional attention mechanisms improve performance metrics (e.g., forecast accuracy) in the context of exchange markets and trading strategies. A specific objective of this study is to evaluate the performance of novel methodologies, including convolutional attention mechanisms, thereby contributing to the development of more robust, data-driven trading systems. In prediction and/or classification tasks addressed by existing literature, \gls{DL} \gls{NN} models possess inherent properties that make them particularly suitable for \gls{MTS} problems. Firstly, these models are robust to noise in input data and can support learning even in the presence of missing values \citep{dixon2015implementing}. Secondly, \gls{DL} \gls{NN} models do not impose strong assumptions about the underlying mapping function, allowing them to absorb both linear and non-linear relationships \citep{huck2009pairs}. This ability is crucial, as most real-life events involve complex, non-linear relationships \citep{huck2010pairs}. Thirdly, \gls{DL} \gls{NN} models exhibit generalization power, enabling them to recognize previously unobserved relationships in data after learning from a given set of inputs \citep{GONCALVES201938}. Fourthly, these models are flexible in their treatment of input data, not enforcing persistence of any specific distribution \citep{dorffner1996neural}. Fifthly, they handle heteroskedasticity more effectively due to their ability to detect hidden relationships without the imposition of additional constraints \citep{gonccalves2023variable}.

Unsurprisingly, a lively strand of research documents that \gls{DL} \gls{NN} architectures have been successfully implemented in various application fields \citep{hatcher2018survey}, and financial exchange markets have recently started to benefit from the learning capability of these models. In general, researchers aim to use \gls{DL} \gls{NN} architectures to predict or classify trends and detect anomalies. \cite{ding2015deep} assesses stock market price predictions by implementing a \gls{DL} \gls{NN} to learn event embeddings and a \gls{CNN} for short-, medium-, and long-term analyses, improving accuracy and profit relative to basic \gls{NN}s. \cite{heaton2016deep} uses a \gls{DL} auto-encoding technique based on \gls{PCA} for dimensionality reduction, allowing for feature extraction and defining a smart index for stocks. Similarly, \cite{korczak2017deep} confirms that using a \gls{CNN} within the H2O algorithmic trading framework significantly improved the average rate of return per transaction in the FOREX market. Additionally, \gls{RNN}s in general and \gls{LSTM} in particular are capable of learning temporal dependencies from context, being widely considered as benchmark models for \gls{MTS} problems \citep{hochreiter1997long}. 

\cite{fischer2018deep} adopt \gls{LSTM} networks for predicting stock price movements, showing that these layers had better predictive power than memory-free classification networks. In turn, \cite{schnaubelt2022deep} presents a \gls{DRL} application to optimize execution in cryptocurrency exchanges by learning optimal limit order placement strategies. Here, a specialized training environment is designed, incorporating a purpose-built reward function, market state features, and a virtual limit order exchange. Using 18 months of high-frequency data from major cryptocurrency exchanges -- covering 300 million trades and 3.5 million order book states -- \gls{DRL} algorithms are compared against benchmarks. This study concludes that proximal policy optimization effectively learns superior order placement strategies, adapting aggressiveness based on execution probabilities influenced by trade and order imbalances. Moreover, \cite{alfonso2024optimizing} apply \gls{DRL} to automate optimal credit card limit adjustment policies. Using historical data from a Latin American super-app, the decision-making problem is framed as an optimization task balancing revenue maximization and provision minimization. An offline learning strategy is employed to train the \gls{DRL} agent, with a Double Q-learning approach demonstrating superior performance over other strategies.

 \cite{kriebel2022credit} explore \gls{DL} \gls{NN} models and other techniques to extract credit-relevant information from user-generated text on Lending Club. Findings indicate that even short pieces of text significantly enhance credit default predictions. An information fusion analysis further confirms the value of textual data. \gls{DL} generally outperforms other text-based approaches, and a comparison of six \gls{DL} \gls{NN} models architectures, including transformer models like BERT and RoBERTa, shows similar performance, suggesting that simpler methods, such as average embedding \gls{NN}s, can be as effective as more complex models for credit scoring. Recently,\cite{zhong2024distributed} propose a distributed mean reversion online portfolio strategy using a stock correlation sub-network to address limitations in existing strategies, such as their lack of universality and the restriction on short selling. A theoretical analysis confirms its generalization and convergence rate. Empirical results demonstrate superior return performance compared to existing universal strategies while maintaining robustness against transaction costs, making it more suitable for real-time investment applications. In the context of exchange markets, \cite{rzayev2025adoption} evaluate the impact of adoption timing on cryptocurrency markets by decomposing total adoption into innovators (i.e., early adopters) and imitators (i.e., late adopters). Findings indicate that innovators drive the relationship between user adoption and cryptocurrency returns, enhancing price efficiency, whereas imitators contribute to price noise. Additionally, the adoption model effectively captures market phenomena such as herding behavior, improving cryptocurrency pricing forecasts. The proposed framework offers a valuable approach for studying market dynamics and can be applied across various domains in financial and operational research.

In addition to establishing a fully integrated end-to-end framework absent in existing literature, this study presents two key findings demonstrating the power and applicability of convolutional attention mechanisms and roll padding in exchange market forecasting and classification tasks. First, results indicate that convolutional attention mechanisms significantly improve forecasting accuracy by effectively capturing complex patterns and dependencies in market data, enhancing prediction performance. Second, findings show that integrating these technical innovations into automated trading systems strengthens decision-making capabilities, leading to more effective and profitable trading strategies in exchange markets.

The study is organized as follows. Section \ref{sec-2} presents the case study, including the trading and software implementation frameworks, as well as data pre-processing steps and feature engineering. Section \ref{sec-3} provides an overview of all \gls{DL} \gls{NN} architectures considered in this study, including the proposed technical innovations -- roll padding and convolutional attention mechanisms. Results are presented in Section \ref{sec-4}. Section \ref{sec-5} compiles managerial implications. Section \ref{sec-6} concludes.

\section{Case Study}\label{sec-2}
\subsection{Betfair Trading Characteristics}
The goal of this research is to accurately estimate changes in odds for buying and selling bets while attempting to guarantee a profit. The analysis focuses on the Betfair betting exchange, the largest platform of its kind globally, with a predominant customer base in the UK \citep{brown2017role}. Modeling exchange markets requires considering a platform where individuals and entities trade fungible items of value with low transaction costs, at prices determined by supply and demand. Extracted models represent multiple interactions among participants. In most modeling scenarios, unforeseen external factors can disrupt assumptions on which predictive systems rely. However, within this restricted time frame, the market under consideration operates as a closed-loop system, where only internal market data influence price fluctuations. Consequently, predictive accuracy depends exclusively on market data, so that this study is solely concerned with purely speculative markets. 

Among real-world exchange markets that adhere to the closed-loop interaction property, betting exchanges serve as a prime example. These facilitate the trading of bookmaking contracts, structured as binary options (i.e., win or lose), where the payoff is either a fixed monetary amount or nothing, depending on the outcome of a future event \citep{Chen:2008:PCM:1374376.1374421,kn:spmining}. Betting exchanges primarily operate within sports markets but also offer trading opportunities on elections and other event-based markets. Analogous to financial markets, buying and selling operations correspond to betting for and against an outcome (i.e., Back and Lay bets). The proposed methodology applies to exchange markets that provide market depth access, also known as level 2 market data. Relevant examples include futures (e.g., Dorman Trading, Phillip Capital), forex (e.g., FXCM), securities (e.g., Euronext Bonds), betting exchanges (e.g., Betfair, Betdaq, Matchbook), and cryptocurrency exchanges (e.g., Coinbase, Bitmex). These markets share the same fundamental structure, enabling adaptation into the presented framework. In contrast, certain exchange types, such as contracts for difference (CFDs), which involve exchange virtualization and do not provide market depth data, cannot be considered. 

Table \ref{table-ladder} presents a snapshot example of a market depth view. This information is referred to throughout the manuscript as a \gls{RDF}. The ``Price'' column represents the ladder of possible transaction prices. The market buy and sell amounts are listed in the ``Bid'' and ``Ask'' columns, respectively. The ``Buy'' and ``Sell'' columns indicate the agent’s own orders awaiting execution. When buy and sell orders converge at the same price, a matched transaction occurs, with the corresponding transacted amount recorded in the ``Volume'' column. The yellow cell highlights the most recent matched price. 

The Betfair exchange is selected as a case study due to its accessible and free \gls{API}, which facilitates the retrieval of raw market data. Bets are bought and sold at varying prices, commonly referred to as odds. The price dynamics enable the realization of \gls{PL} before the final event outcome is known. Depending on market sentiment, price movements can span a few or several ticks,\footnote{A tick denotes the smallest unit of a price change.} making price volatility a widely acknowledged characteristic. The UK To-Win horse racing market is distinguished by high liquidity and volatility levels, factors that are critical for aligning with the research objectives. In this study, market engagement is restricted to the 10-minute pre-race period. This time frame is strategically selected because, before a race begins, each runner's price is predominantly influenced by speculative activity. Market movements during this period are driven solely by internal trading data, establishing a strongly closed-loop system. Given this atomistic property, the analysis focuses exclusively on purely speculative markets.  

\begin{table}[htbp]
	\footnotesize
	\begin{center}
		\begin{tabular}{|c||c|c|c||c||c|}
			\hline
			Buy&Bid&Price&Ask&Sell&Volume\\
			\hline
			\hline
			&    &\cellcolor{mygreen} \rvdots&\cellcolor{mylightblue}    &    & \cellcolor{gray!30}   \\
			\hline
			&    &\cellcolor{mygreen}5,1& \cellcolor{mylightblue}   &    & \cellcolor{gray!40}20\\
			\hline
			&    &\cellcolor{mygreen}5,0&\cellcolor{myblue}250&    &\cellcolor{gray!40}93\\
			\hline
			&    &\cellcolor{mygreen}4,9&\cellcolor{mylightblue}    &    &\cellcolor{gray!40}68\\
			\hline
			&    &\cellcolor{mygreen}4,8&\cellcolor{myblue}263&    &\cellcolor{gray!40}24\\
			\hline
			&    &\cellcolor{mygreen}4,7&\cellcolor{myblue}148& \cellcolor{myblue}10,00&\cellcolor{gray!40}70\\
			\hline
			&    &\cellcolor{yellow}4,6&\cellcolor{myblue}349&\cellcolor{myblue}5,00&\cellcolor{gray!40}76\\
			\hline
			&\cellcolor{mypink}8&\cellcolor{mygreen}4,5&   &    &\cellcolor{gray!40}217\\
			\hline
			&\cellcolor{mypink}2&\cellcolor{mygreen}4,4&    &    &\cellcolor{gray!40}23\\
			\hline
			\cellcolor{mypink}10,00&\cellcolor{mypink}10&\cellcolor{mygreen}4,3&    &    & \cellcolor{gray!40}4\\
			\hline
			&\cellcolor{mypink}448&\cellcolor{mygreen}4,2&    &    &    \\
			\hline
			&\cellcolor{mypink}398&\cellcolor{mygreen}4,1&    &    &    \\
			\hline
			&\cellcolor{mypink}335&\cellcolor{mygreen}4,0&    &    &    \\
			\hline
			& \cellcolor{mylightpink} &\cellcolor{mygreen}\rvdots&    &    &    \\
			\hline
		\end{tabular}
	\end{center}
	\caption{Snapshot of market depth RDT information.}
	\label{table-ladder}
\end{table}

\begin{figure}[H]
	\centering
	\begin{subfigure}{0.32\textwidth}
		\centering
		\includegraphics[width = \textwidth]{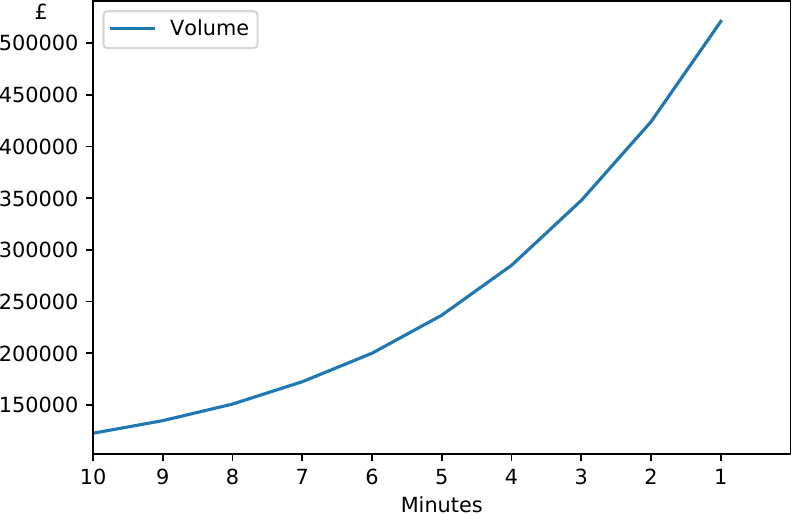}
		\label{fig:right}
	\end{subfigure}
	\begin{subfigure}{0.32\textwidth}
		\centering
		\includegraphics[width = \textwidth]{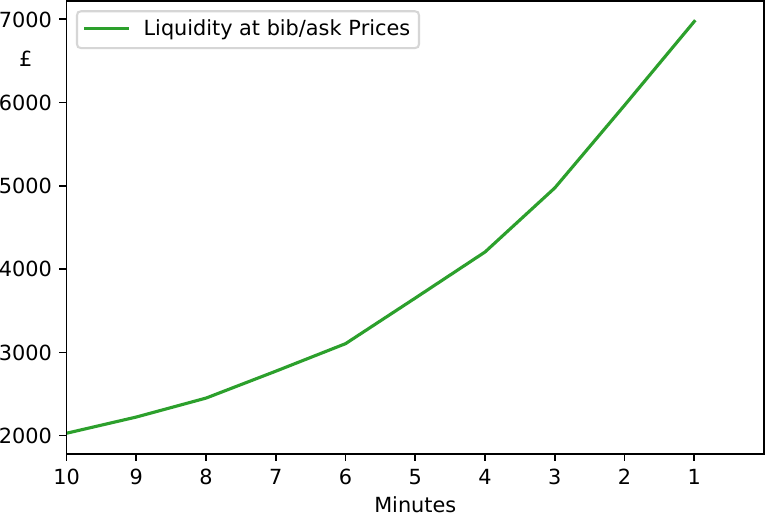}
		\label{fig:left}
	\end{subfigure}
	\begin{subfigure}{0.32\textwidth}
		\centering
		\includegraphics[width = \textwidth]{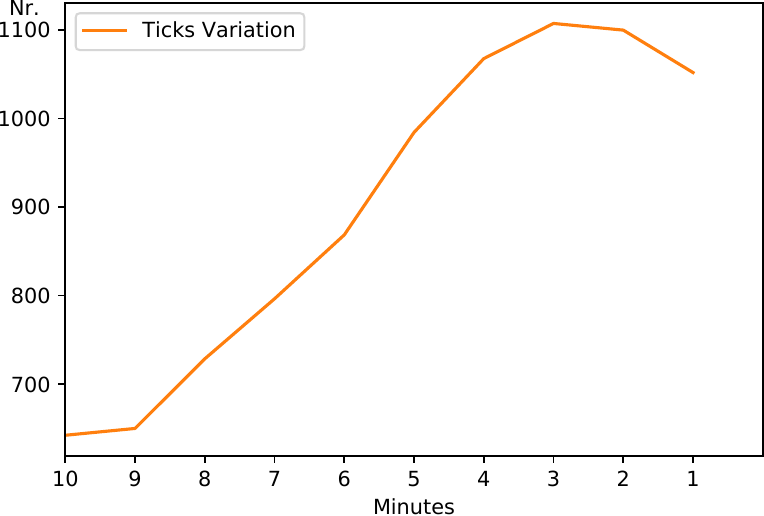}
		\label{fig:right}
	\end{subfigure}
	\caption{Average trading volume, average liquidity at the bid and ask price, and average number of ticks variation in absolute value per minute.}
	\label{fig-pre-live-data}
\end{figure}

Fig. \ref{fig-pre-live-data} illustrates the average trading volume, liquidity (i.e., the total sum of unmatched amounts at the bid and ask prices), and volatility (i.e., the absolute number of tick variations per minute) across the observed sample of races, including all participating runners. From the tenth minute before race commencement to the race's official start, average trading volume increases by approximately 4.2 times, average liquidity rises by about 3.4 times, and average volatility escalates by nearly 1.6 times. From a dynamic perspective, this means that all key market variables exhibit an upward trend as the race start approaches.

Raw data were collected directly from Betfair servers in real-time at a rate of two frames per second from September 1, 2014, to August 29, 2016. For conciseness, summary statistics on trading volume, liquidity, and volatility at the race level, analyzed on a per-minute basis within the 10-minute window before each race, are provided in Table \ref{table-stats-sum}. The dataset comprises a customized collection of UK To-Win horse racing markets, encompassing 15 to 30 daily races, with each race featuring between 3 and 25 competing horses. Differently from \cite{GONCALVES201938}, market data is used to predict price variation for only the two final minutes before a race begins and not continuously throughout the 10 minutes pre-live. This form of trading, characterized by short market exposure and a strong reliance on real-time market data, aligns with day trading principles. When executed automatically over very short timeframes, such trading strategies fall under high-frequency trading (HFT). The methodologies implemented in this study belong to these trading categories. Access to market depth data is crucial in both high-frequency trading and day trading. Unlike long-term trading, where broader trends are considered, short-term trading requires detailed real-time information on price movements at multiple levels. To obtain this data, traders -- or automated trading agents -- rely on a ladder view of the market, as illustrated in Table \ref{table-ladder}. All price forecasting data used in this study are derived from the continuously evolving price ladder.

\begin{table}[!htb]
	\begin{center}
				\footnotesize
		\begin{tabular}[t]{ccccc}
			\hline \addlinespace[0.5ex]
			& Mean & Std Dev & Min & Max  \\ \hline \addlinespace[0.5ex]
			Volume & & & & \\ \hline \addlinespace[0.5ex]
			1st min & 520865 & 275541 & 14242& 4858654 \\  
			
			2nd min	& 423930 & 247598 &	10783 &	4713870 \\  
			
			3rd min	& 347772 & 223834 &	8390 &	4569493 \\  
			
			4th min	& 285066 & 201834 & 7014 & 4237354 \\  
			
			5th min	& 236840 & 183575 & 5818 & 3944767 \\  
			
			6th min	& 200076 & 168577 & 4514 &	3779854 \\  
			
			7th min & 172516 & 156206 &	2524 &	3554934 \\  
			
			8th min & 150905 & 145219 & 1898 & 3481841\\  
			
			9th min & 134880 & 137576 &	1647 & 3450423\\  
			
			10th min & 122693 & 131643 & 1503 & 3420282\\ \hline \addlinespace[0.5ex]
			
			Liquidity & & & & \\ \hline \addlinespace[0.5ex]
			1st min &	6974 & 	12301 &	531 & 451241 \\  
			
			2nd min	& 5967	&13151&	682&	468516 \\  
			
			3rd min	& 4975	& 12832	&526&	454538\\  
			
			4th min	& 4205	& 12544	& 393	&453012 \\  
			
			5th min	& 3652	& 12143	& 329	& 428307\\  
			
			6th min	& 3105	& 11137	& 285	& 399753 \\  
			
			7th min & 2774	& 10808	& 245	& 359948 \\  
			
			8th min & 2451	& 9825	& 212	& 331043 \\  
			
			9th min & 2224	& 9540	& 191	& 323377 \\  
			
			10th min & 2029	&8935	&171&	292921\\ \hline \addlinespace[0.5ex]
			
			Volatility$^+$ & & & & \\ \hline \addlinespace[0.5ex]
			1st min &	1052 & 1443 & 77 & 29899\\  
			
			2nd min	& 1100	& 1593	& 0 & 31323 \\  
			
			3rd min & 1107	& 1357	& 22 & 37893 \\  
			
			4th min	& 1067	& 1524	& 0	& 29442 \\  
			
			5th min	& 984	& 1412	& 20 & 33069 \\  
			
			6th min	& 868	& 1437	& 29 & 49078 \\  
			
			7th min & 796	& 1191	& 4	& 30815 \\  
			
			8th min & 729	& 1162	& 7	& 27630 \\  
			
			9th min & 650	& 961	& 2 & 29686\\  
			
			10th min &642	& 1326	& 1	& 38439\\  
			\hline  \addlinespace[-0.5ex]
		\end{tabular}
	\end{center}	
	\caption{Pre-live betfair horse racing markets summary statistics.}
	\label{table-stats-sum}
	\scriptsize\textsuperscript{}Note: Total number of observations (i.e. races): 14421. Each line represents the $x$ minute before the start of a race, $x = \{1st, ..., 10th\}$. Units of measure are clarified in Fig.\ref{fig-pre-live-data} . Symbol $^+$ represents ticks variation in absolute value per minute.
	
\end{table}

To achieve \gls{PL}, price movements in either direction are necessary. Price fluctuations follow the fundamental law of supply and demand. In financial markets, stock prices rise when demand for a company's shares increases and fall when sellers outnumber buyers. The same mechanism applies to betting exchanges. If the majority of participants place Lay bets against a runner (e.g., a horse in a race or a football team), the fixed odds for winning the event will increase. Conversely, if bettors invest in favor of a runner's victory, the odds will decline. The extent of these price movements, measured in ticks, depends on market pressure and investor sentiment, leading to either minor fluctuations or substantial shifts in odds.

\subsection{Trading Implementation Framework}\label{intro:problem}
Betting markets are short-lived, yield easily quantifiable final payoffs for traded assets, and exhibit a degree of repetition, making them well-suited for efficiency testing. In betting exchanges, each market corresponds to a specific event, such as a tennis match or a horse race. Within an event, there are multiple runners (e.g., horses in a horse race), on which bets are placed. Two types of bets exist in betting exchanges: Back and Lay. A Back bet supports a runner to win, whereas a Lay bet wagers on the runner to lose. Bets are placed at specific prices, which correspond to implied probabilities. For example, a price of $2.0$ represents a 50\% probability of winning ($1/2 = 0.5$), a price of $1.01$ implies a 99\% probability ($1/1.01 = 0.99$), and a price of $100$ corresponds to a 1\% probability ($1/100 = 0.01$). Table \ref{table-ladder} presents an example of unmatched bets placed at various price levels.
\begin{itemize}
	\item  Lay of $\pounds{10.00}$ at $4.30$ (Lay $10@4.3$);
	\item  Back of $\pounds{10.00}$ at $4.70$ (Back $10@4.7$); and
	\item  Back of $\pounds{5.00}$ at $4.60$ (Back $5@4.6$).
\end{itemize}

\noindent If a bet is placed at a price that the market is willing to buy, it will be matched at the best available price. For instance, based on the market state in Table \ref{table-ladder}, placing a Back bet of 15@4.4 (on the blue side) will result in a match of \( \pounds 8.00 \) at 4.5 and \( \pounds 2.00 \) at 4.4, leaving the remaining \( \pounds 5.00 \) at 4.4 unmatched on the ask side, awaiting a Lay bet from another participant. The traded volume information will be updated accordingly. This mechanism determines how prices fluctuate in the market. Since this bet was matched at two different prices (\( N=2 \)), the global matched price of the bet can be calculated using the Eq.~(\ref{eq:oddavg}):
\begin{align}\label{eq:oddavg}
{\rm Price\text{ }Average}= \frac{ \sideset{}{_{n=1}^N}\sum {\rm (Price_n \times Amount_n)}}{\sideset{}{_{n=1}^N}\sum {\rm Amount_n}} 
\end{align}

\noindent If a Back bet is placed above the best available offer in the market (e.g., 4.5 in Table \ref{table-ladder}), such as 15@4.9, it will remain unmatched and will be queued in a \gls{FIFO} order with other Back bets at that price, waiting to be matched. Similarly, a Lay bet placed below the best available offer (i.e., a counter bet waiting to be matched) will also remain in the market until a matching Back bet is placed. Only unmatched or partially unmatched portions of bets can be canceled. The profit of a Back bet is calculated using Eq. (\ref{eq:profitBack}), while the liability (i.e., the potential loss) of a Back bet is equal to the amount staked:
\begin{align}\label{eq:profitBack}
{\rm 
Profit\text{ }Back\text{ }Bet = Amount\text{ }Back \times ( Price\text{ }Back - 1)
}
\end{align}

\noindent The liability or amount at risk in the event of a loss for a Lay bet is given by Eq. (\ref{eq:liabilityLay}), while the profit of a Lay bet is equal to the amount of the bet itself. In summary, a Lay bet is the mirror image of a Back bet:
\begin{align}\label{eq:liabilityLay}
{\rm 
Liability\text{ }Lay\text{ }Bet = Amount\text{ }Lay \times ( Price\text{ }Lay - 1)
}
\end{align}

\noindent Using combinations of Back and Lay bets, it is possible to secure a fixed \gls{PL} before the final outcome of an event. An example of such a trade, where the event's result does not need to be known to secure a fixed \gls{PL}, is as follows:
\begin{itemize}
	\item  Back of \pounds{2.00} at 2.12 (Back 2@2.12) Matched; and
	\item  Lay of \pounds{2.00} at 2.10 (Lay 2@2.1) Matched;
\end{itemize}

\noindent For a bet to be matched, it must become the best available offer in the market and be countered by an opposing bet. When the runner wins, the profit from the Back bet minus the loss from the Lay bet is calculated as follows:
\[
2 \times (2.12 - 1) - 2 \times (2.10 - 1) = 2.24 - 2.20 = 0.04
\]
\noindent When the runner loses, the profit from the Lay bet minus the loss from the Back bet is: $2 - 2 = 0$. It is important to note that if a Back and Lay bet combination is placed with the same amount at different prices, a profit will occur if the Back price is higher than the Lay price, or a loss will occur if the Back price is lower than the Lay price, but only if the specified runner wins the event. If any other runner wins, and the amounts for the Back and Lay bets are equal, the \gls{PL} will be zero. The agent can distribute the guaranteed \gls{PL} from one runner across all other runners. To ensure an equal distribution of the \gls{PL} across all possible outcomes, the amount required to close the trade must be recalculated. This process is known as \textit{greening} or \textit{hedging}. If a Back position is open in the market, the amount needed to close the position with the corresponding Lay bet is calculated using Eq. (\ref{eq:amountLaycalc}):
\begin{align}\label{eq:amountLaycalc}
{\rm 
Close\text{ }Amount\text{ }Lay = \frac{Price\text{ }Open\text{ }in\text{ }Back}{Price\text{ }Lay\text{ }to\text{ }Close} \times Amount\text{ }Open\text{ }in\text{ }Back
}
\end{align}
If a Lay position is open on the market, the amount to close the position with the corresponded Back is calculated using Eq. (\ref{eq:amountback}).
\begin{align}\label{eq:amountback}
{\rm 
Close\text{ }Amount\text{ }Back = \frac{Price\text{ }Open\text{ }in\text{ }Lay}{Price\text{ }Back\text{ }to\text{ }Close} \times Amount\text{ }Open\text{ }in\text{ }Lay
}
\end{align}

\subsection{Software Implementation Framework}\label{sec-software-framework}
\begin{figure}[H]
	\begin{center}
		\leavevmode
		\includegraphics[width=0.6\textwidth]{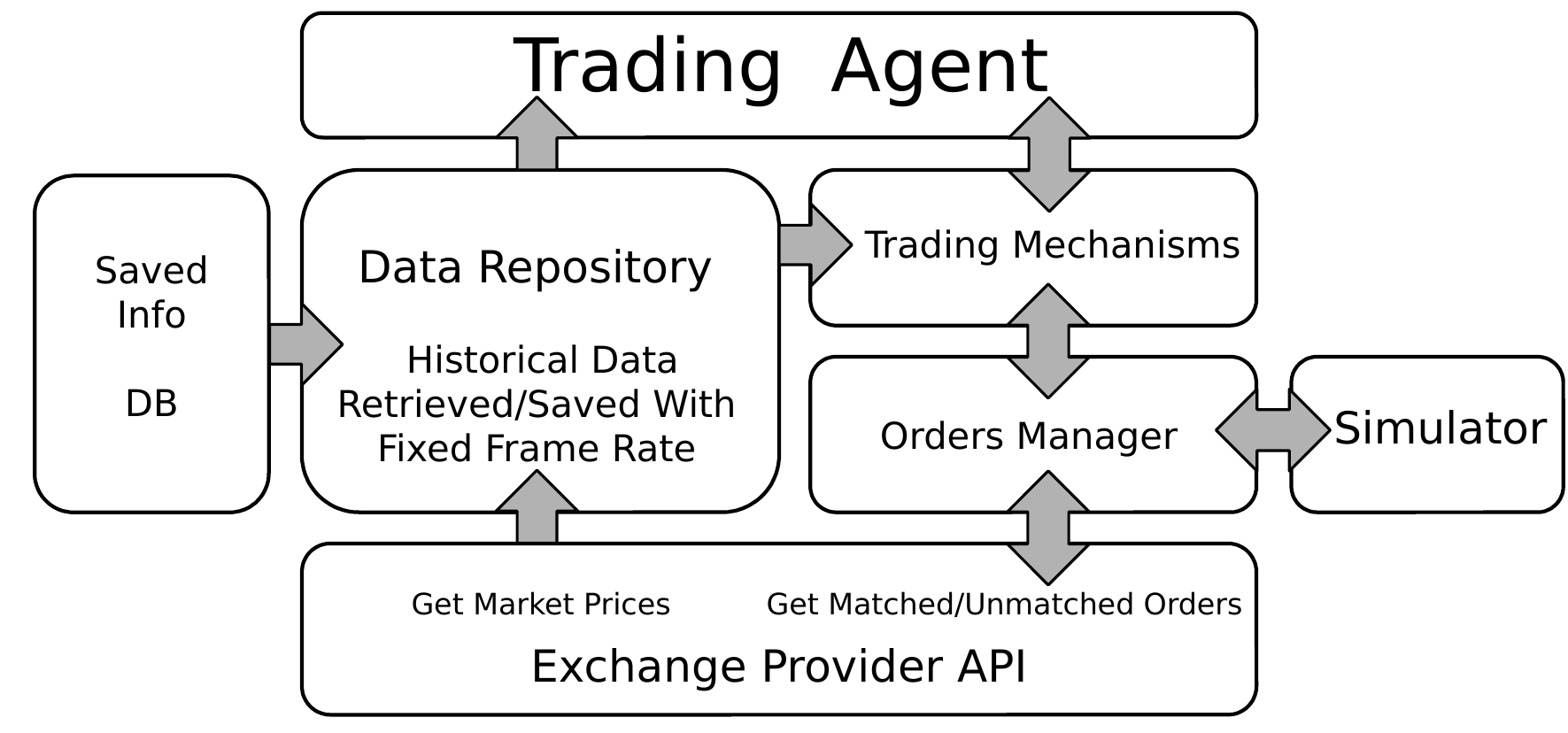}
		\caption{High level architecture for automated betting exchange.}
		\label{fig-trade-architecture}
	\end{center}
\end{figure}

The developed software is based on an event-driven architecture \citep{5711108,kn:reactiveagents,kn:reusablepatterns}, where modules communicate through interfaces designed around architectural patterns that facilitate the production, detection, consumption, and reaction to events. Fig. \ref{fig-trade-architecture} shows the primary modules and their interconnections. Parallel-processing techniques \citep{289940,yau1995architecture} are also incorporated, allowing for the instantiation of multiple Trading Agents, each with distinct policies, running concurrently and managing multiple trades simultaneously. This software framework we developed for automatic trading is called JBet\footnote{The JBet trading software framework source code can be consulted at \url{https://github.com/rjpg/JBet}.}.

Next, a description of each module is provided.

\begin{description}
	\item [$\bullet$] \textbf{Exchange provider API:} This module serves as the interface between the trading system and the external betting exchange platform. It handles communication protocols and facilitates the retrieval and submission of data, such as market information, odds, and transaction details. The Exchange API ensures that the system can access real-time market data and execute trades in the betting exchange. The Betfair exchange \gls{API} is used to access Betfair data \citep{betfair-api,pitt2005betfair}, where Betfair acts as a service provider and other companies are clients. Its primary focus is speed and manageability, commonly employed in the development of trading software or software for tipsters. It can also be used to develop autonomous agents. In this framework, the Betfair \gls{API} operates as a low-level communication interface between the framework and the exchange server. This layer interacts with the \textit{Data Repository} module to provide data regarding the prices and volumes of each runner, and with the \textit{Orders Manager} module to supply information on the states of bets and manage the placing and canceling of bets. According to \cite{betfair-api}, main Betfair \gls{API} services  used for these actions include:\footnote{Depending on the type of licensing, a different number of calls per minute are permitted for each service.}
\end{description}

\begin{itemize}
	\item \textit{Data Repository}\vspace{-.2cm}
	\begin{itemize}
		\item  Get Complete Market Prices; and
		\item  Get Market Traded Volume.
	\end{itemize}
	\item \textit{Orders Manager}\vspace{-.2cm}
	\begin{itemize}
		\item  Get Matched and Unmatched Bets;
		\item  Place Bets;
		\item  Update Bets; and
		\item  Cancel Bets.
	\end{itemize}
\end{itemize}

\begin{description}
	\item [$\bullet$] \textbf{Data Repository:} This module is responsible for gathering data, notifying listeners about new data or changes in the market state, saving data, and enabling the replay of saved data. The main interface of this module is shown in Listing \ref{data-interface}. The \textit{Market Update} event type simply notifies listeners that new data regarding the prices and volumes of the runners has arrived. Trading can occur both before and after the start of a race, which is reflected in the \textit{Market Live} event type, activated when the market becomes in-play. The \textit{Market Suspended} event type is triggered when the market is suspended, which may occur for various reasons depending on the type of event. For example, in a soccer match, the market is suspended after a goal until play resumes. Furthermore, markets are briefly suspended right after they go in-play. The \textit{Data Repository} module can be instantiated with a new market by an external object, and when this occurs, the \textit{Market New} event is triggered.\footnote{This module can be connected to the Betfair server via the Betfair \gls{API} or through saved files, which can be used for data replay and simulation, as further explained below in this section.}
\end{description}

\vspace{-.2cm}

\lstset{ %
	language=JAVA,                
	basicstyle=\scriptsize,       
	numbers=left,                   
	numberstyle=\scriptsize,      
	frame=single,
}

\begin{small}
\begin{lstlisting}[language=JAVA,caption={Market Change Listener Interface},label=data-interface]
public interface MarketChangeListener {
	
	public enum EventType {MarketUpdate, MarketLive, MarketClose, MarketSuspended, MarketNew}
	
	public void MarketChange(MarketData md, MarketChangeListener.EventType marketEventType);
}
\end{lstlisting}
\end{small}

\begin{figure}[htbp]
	\begin{center}
		\leavevmode
		\includegraphics[width=0.9\textwidth]{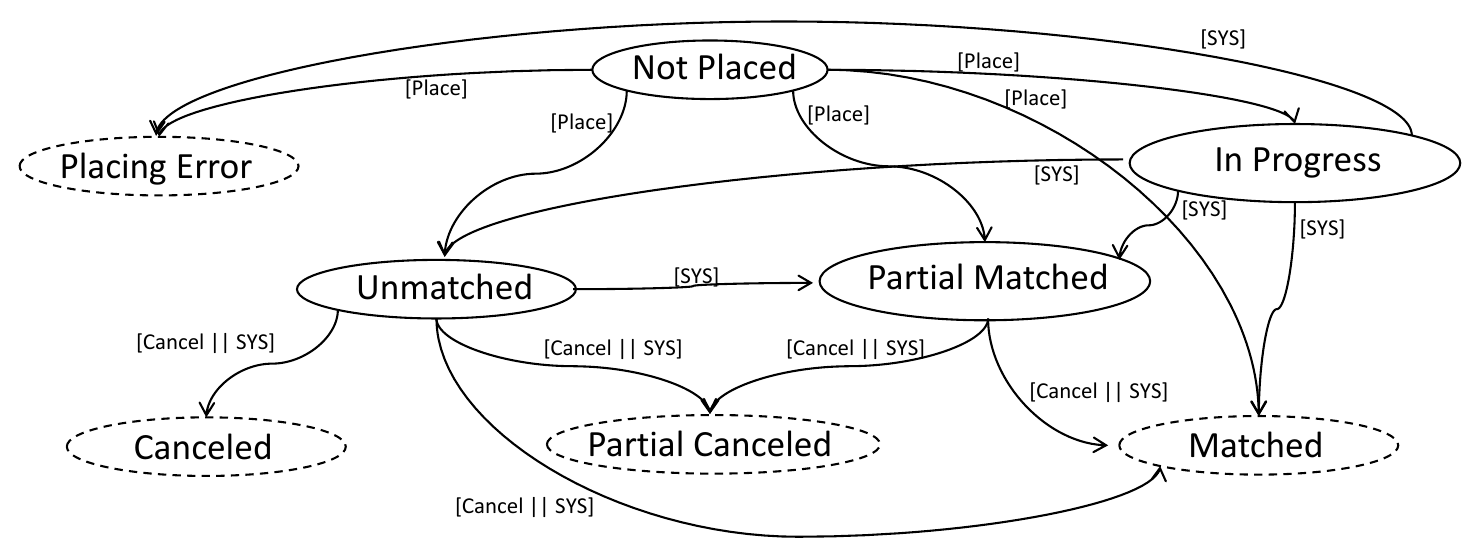}
		\caption{State machine for a given order.}
		\label{fig-bet-state}
	\end{center}
\end{figure}

\begin{description}
	\item [$\bullet$] \textbf{Orders Manager:} This module ensures the proper handling of bet information and manages all objects with a \textit{BetListener} interface (i.e., \textit{Trading Mechanisms}), notifying them about the state of their bets. Centralizing all bet processing is crucial to optimize the number of calls to the \textit{Get Matched and Unmatched Bets} service, which is subject to limits. Additionally, in some cases, the Betfair \gls{API} does not return the ID of a placed bet, leaving the program uncertain about the bet's placement. In such instances, the \textit{Bets Manager} module tracks the bets without owner and correctly reassigns them (cf. Fig. \ref{fig-bet-state}, state \textit{In Progress}). Fig. \ref{fig-bet-state} presents the possible states and transitions of a bet within the framework. The $[Place]$ and $[Cancel]$ transitions are initiated by the agent or \textit{Trading Mechanism} modules (i.e., client-side active actions). The $[SYS]$ transitions are handled by the system, such as when the system automatically changes the state from \textit{unmatched} to \textit{matched} once the order is filled. This state machine serves as the foundational structure, extensively used by the \textit{Trading Mechanisms} module to manage the market agent's position. 
\end{description}

\begin{description}
	\item [$\bullet$] \textbf{Trading Agents:} To instantiate a \textit{Trading Agent} object, it must be extended from the abstract superclass \textit{Bot}. This class implements all the interfaces and virtual methods necessary to interact with the \textit{Data Repository} and \textit{Trading Mechanisms} modules. \textit{Trading Agent} objects are typically attached to a single market, observing a single runner, but they can also observe multiple markets and runners simultaneously, which is useful for techniques such as \textit{dutching} and \textit{bookmaking}. The \textit{Trading Agent} object can also initialize \textit{Trading Mechanisms} objects (i.e., trading processes) whenever it forms a conclusion about a runner’s forecast. When a \textit{Trading Mechanism} starts, it runs in parallel, and the \textit{Trading Agent} is continuously informed about the state of the trade throughout its execution. At this higher abstraction level, it becomes easier to implement decision policies for interacting with the markets, ranging from simple rule-based decision policies to more complex methodologies, such as time series predictions. 
\end{description} 

\begin{description}
	\item [$\bullet$] \textbf{Trading Mechanisms:} This module is used to discipline the trader's attitude towards the market. In other words, these methods are designed to be implemented on a computer. They are executed after a decision regarding the market forecasting is made. Once the decision for the depreciation or appreciation of a runner is taken, a sequence of steps is initiated to maximize profit. This study focuses on three trading methodologies:
\end{description}
\begin{enumerate}
	\item  Scalping;
	\item  Swing; and
	\item  Trailing-Stop.
\end{enumerate}

\noindent In the framework, these methods are implemented within the \textit{Trading Mechanism} module (cf. Fig.~\ref{fig-trade-architecture}). After a \textit{Trading Agent} parametrizes and instantiates a \textit{Trading Mechanism}, it runs in parallel and continuously informs the owning agent about the state of the trade. Ultimately, it will notify the agent when the trade is complete and provide the operation's \gls{PL}.

\begin{figure}[htbp]
	\begin{center}
		\leavevmode
		\includegraphics[width=0.7\textwidth]{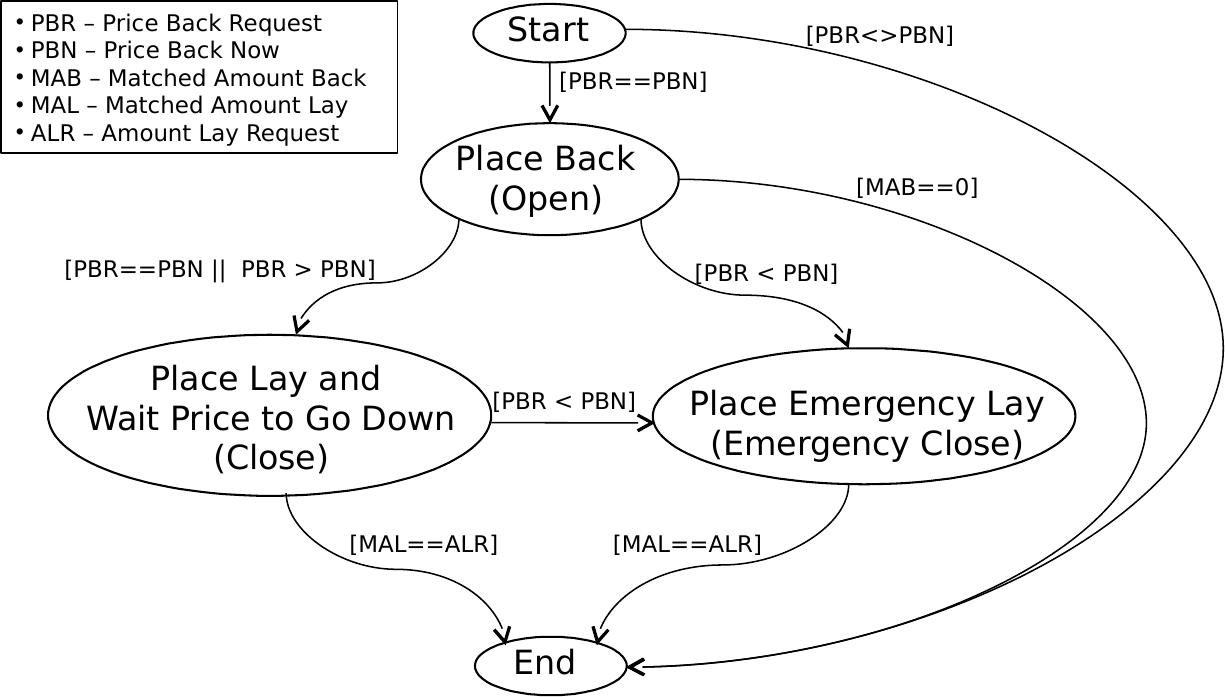}
		\caption{Simplified graph scheme for a Back$\Rightarrow$Lay scalp implementation.}
		\label{fig-scalp}
	\end{center}
\end{figure}

\begin{description}
	\item [$\bullet$]{\textbf{Scalping:}} It is a strategy used for very short-term trading, where the trader aims to make multiple small profits over time. These small profits accumulate, leading to a significant total gain. Scalping is most effective in markets with high liquidity and many active participants. It works particularly well in markets like Betfair, where the trader can take advantage of price fluctuations within short timeframes. The concept of scalping is simple: if a Back bet is placed at a certain price, a Lay bet must be placed at the next lower price, or vice versa for the opposite direction. The \gls{PL} is determined by the difference -- or spread -- between the Back and Lay prices, as explained in Subsection \ref{intro:problem}. Betfair's betting exchange is ideal for this type of trading, especially in markets like horse racing, where liquidity is high, particularly just before the start of a race. Scalping involves trading in the market tick by tick, where one tick is a single step in the price scale of the ladder. For example, if a Back bet is placed at 2.12, one successful scalp would close the position with a Lay bet at 2.10 (i.e., one tick down). If a trade begins with a Back bet, it suggests that the price is expected to go down. Conversely, if a price is predicted to rise, the scalp begins with a Lay bet.	Fig. \ref{fig-scalp} represents the state machine used to process a Back$\Rightarrow$Lay scalping strategy (i.e., predicting the price to drop). A Lay$\Rightarrow$Back scalp follows a mirror of this state machine. The Price Back Request (\texttt{PBR}) represents the price at which the agent enters the market, while the Price Back Now (\texttt{PBN}) is the current market price. If the price has already moved (i.e., [\texttt{PBR} $\neq$ \texttt{PBN}]) when the order reaches the scalp module (i.e., the starting state), it is assumed that the opportunity has been lost (i.e., the prices have already dropped) or the prices have moved in the wrong predicted direction. In this case, the process ends without further action. If [\texttt{PBR} $==$ \texttt{PBN}], the position is opened in the market with a Back bet. After placing the bet, if it is not matched within a certain timeframe, the trade will end by canceling the bet, assuming that the price has already moved in the predicted direction, and the opportunity is lost. Otherwise, the system will try to close the position by placing a Lay bet. If the price moves down by one tick, the trade will close with a profit. If the price remains the same, the system will wait for a specified period. If the close bet has still not been matched after this waiting period, the system will attempt to close at the same price, resulting in a null profit/loss. If the price rises, the position will close ``in emergency'' with a loss. Listing \ref{method:scalping} presents the declaration of the constructor method for the object that runs the scalping process in parallel. The \texttt{MarketData md} argument identifies the event (e.g., horse race, soccer game, tennis match), while the \texttt{RunnersData rd} argument specifies the runner to be traded. The \texttt{double entryAmount} represents the initial stake of the trade, and the \texttt{double entryPrice} is the price at which the position is opened in the market. The \texttt{int waitFramesNormal} defines the number of updates received from the \textit{Data Repository} before attempting to close the position at the same price it was entered (i.e., with a zero profit/loss). Once the \texttt{int waitFramesNormal} period expires, the \texttt{int waitFramesEmergency} countdown begins. After it expires, the system will close the position at the best available offer to place the counter bet (i.e., an ``emergency'' close with a loss). The \texttt{Bot botOwner} is the agent that owns the trade and is used to keep track of the scalp's state. Finally, the \texttt{int direction} argument indicates the predicted direction of the price movement.	
\end{description}

\begin{lstlisting}[language=JAVA,caption={Main parameters for Scalping mechanism },label=method:scalping]
public Scalping(MarketData md, 
                RunnersData rd,  
                double entryAmount, 
                double entryPrice, 
                int waitFramesNormal, 
                int waitFramesEmergency, 
                Bot botOwner, int direction, ...);
\end{lstlisting}
\begin{description}
	\item [$\bullet$]{\textbf{Swing:}} This trading strategy is very similar to scalping, with the main difference being the number of ticks the price must move before entering the close state \citep{Carter:2007:MTP:1208007}. In the swing methodology, it is possible to define an offset number of ticks for both profit and loss. If the price remains within this offset interval, no action is taken. A swing with an offset of 1 tick for profit and 1 tick for loss is equivalent to scalping. Listing \ref{method:swing} describes the constructor for initializing the swing process. Besides the same parameters present in the scalping constructor, there are additional parameters: \texttt{int ticksUp} and \texttt{int ticksDown}, which represent the offset number of ticks to close the position in profit and loss, depending on the \texttt{direction} parameter. There are also two new arguments: \texttt{boolean frontLine} and \texttt{int waitFramesOpen}. These parameters are used when the agent does not want to enter the market immediately but prefers to wait until the market reaches the price specified in the \texttt{entryPrice} argument. If \texttt{waitFramesOpen} expires and the market does not match the entry bet, the trade process is canceled. If \texttt{frontLine = true}, the agent ignores the \texttt{waitFramesOpen} time and assumes that the agent wants to enter the market at the \texttt{entryPrice} where the counter offer is available.
\end{description}

\begin{lstlisting}[language=JAVA,caption={Swing constructor example},label=method:swing]
public Swing(MarketData Market,
                RunnersData rd,
                double entryAmount,
                double entryPrice,
                boolean frontLine,
                int waitFramesOpen,
                int waitFramesNormal,
                int waitFramesEmergency,
                Bot botOwner,
                int direction,
                int ticksUp,
                int ticksDown);
\end{lstlisting}

\begin{figure}[htbp]
	\begin{center}
		\leavevmode
		\includegraphics[width=0.7\textwidth]{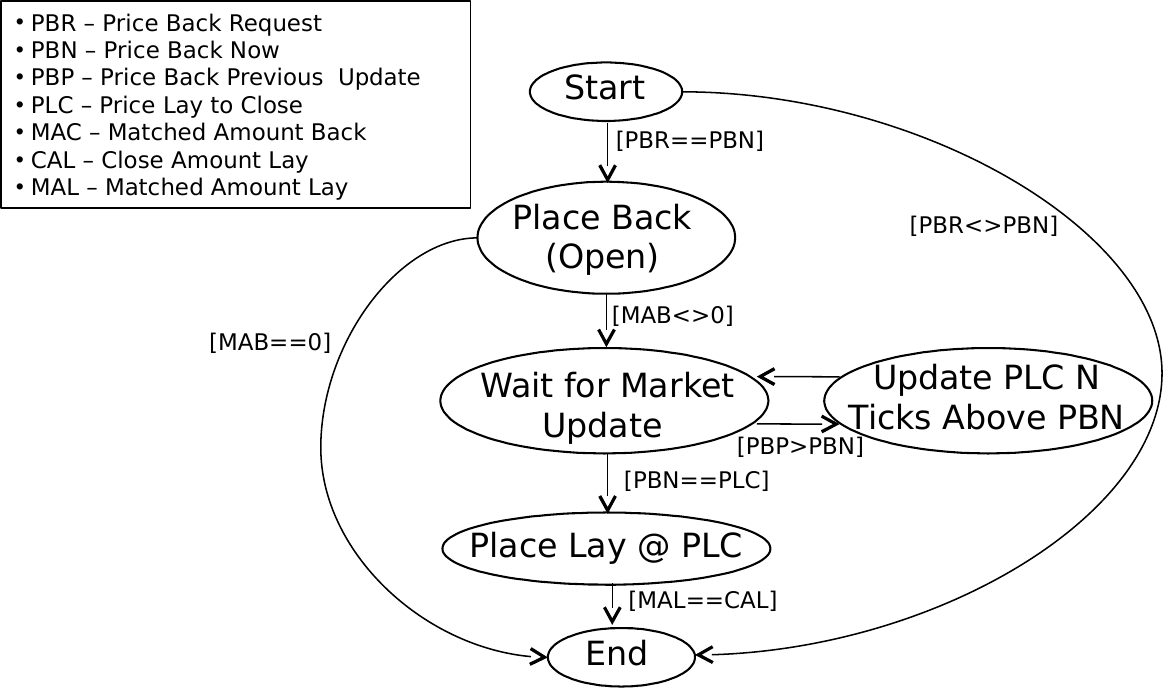}
		\caption{Simplified graph scheme for a Back$\Rightarrow$Lay Trailing-Stop implementation.}
		\label{fig-Trailing}
	\end{center}
\end{figure}

\begin{description}
	\item [$\bullet$]{\textbf{Trailing-Stop:}} This methodology is used when the agent is aiming to capture a broader trend in a market but still wants to maintain a stop-loss condition if the trend reverses. The concept is straightforward: after a position is opened in the market, the close bet is set to close with a tick offset behind the current price and moves only when the price moves in the predicted direction. Eventually, the price will reverse, reaching the close price, at which point the order is placed to close the trade.	Fig. \ref{fig-Trailing} represents the state machine used to process this methodology when predicting a price drop (i.e., Back$\Rightarrow$Lay). The \texttt{Price Lay to Close} (\texttt{PLC}) represents the dynamic price, N ticks above the \texttt{PBN}. The state \texttt{Update PLC N Ticks Above PBN} repeatedly updates the price when [\texttt{PBP} $>$ \texttt{PBN}] (i.e., when the runner price moves in the predicted direction -- down in this case). The \texttt{PLC} is updated only when the \texttt{PBN} moves in the predicted direction. When the price reverses, it eventually reaches the \texttt{PLC}. When this happens, the close order (i.e., Lay bet) is placed. Finally, when [\texttt{MAL} $=$ \texttt{CAL}], it means the close bet is fully matched, indicating that the price has moved in the reverse direction (i.e., up), filling the close bet completely and closing the trade. Listing \ref{method:trainlingstop} presents the constructor method of the object that runs the trailing-stop process in parallel. The \texttt{offset} represents the number of ticks by which the trailing stop follows the runner's price. Additionally, the \texttt{waitFramesNormal} parameter is included to control the duration of the trade, specifying the number of frames (or updates) after which the trade will close, should the price not move in the predicted direction within that time.
	
\end{description}

\begin{lstlisting}[language=JAVA,caption={Trailing-Stop constructor example},label=method:trainlingstop]
public TrainlingStop(MarketData Market,
                        RunnersData rd,
                        double stakeSize,
                        double entryPrice,
                        boolean frontLine,
                        int waitFramesOpen,
                        int waitFramesNormal,
                        int waitFramesEmergency,
                        Bot botOwner,
                        int direction,
                        int offset);
\end{lstlisting}
\begin{description}
	\item [$\bullet$] \textbf{Simulation:} This module addresses limitations involved in simulating markets of this type. The process involves the simulation of bet placement. There are two main problems when simulating a bet placement on the market:
	
	\begin{description}[noitemsep]
		\item [(I)] The first major issue is the influence of the bet amount on the market. Unmatched bets do not appear in the real market, and matched bets do not consume or alter the available amounts in the real market. This issue is unavoidable because, in simulation, the actual amount of the bets is not placed (e.g., this limitation makes it impossible to simulate and test \textit{Trading Agents} relying on \textit{spoofing} methodologies).
		\item [(II)] The second problem is the simulation of the matching process. Bets of all traders in the market are placed in a \textit{FIFO} queue for each price on the runner. It is impossible -- since there is no data provided by the Betfair \gls{API} -- to know in which position our bet is in the queue. It is possible to make an approximate guess by observing the volume matched at the placement price and monitoring its evolution. However, due to the high-frequency nature of these markets, it is impossible to know the exact volume at the price when the placement order reaches the Betfair server. Moreover, it is impossible to determine if canceled bets were placed ahead or behind our bet, which compromises the volume monitoring approach to resolve this issue. For the framework described, we assume the worst-case scenario: bets are considered to be at the front of the queue when the volume transacted is equal to the amount that was in front of the order when it was placed. After that, we monitor the amount matched by tracking the volume variation. An order is considered fully matched when the volume variation reaches the order amount.
	\end{description}
\end{description}

\subsection{Data Collection and Feature Engineering}
\begin{figure}[htbp]
	\begin{center}
		\leavevmode
		\includegraphics[width=0.7\textwidth]{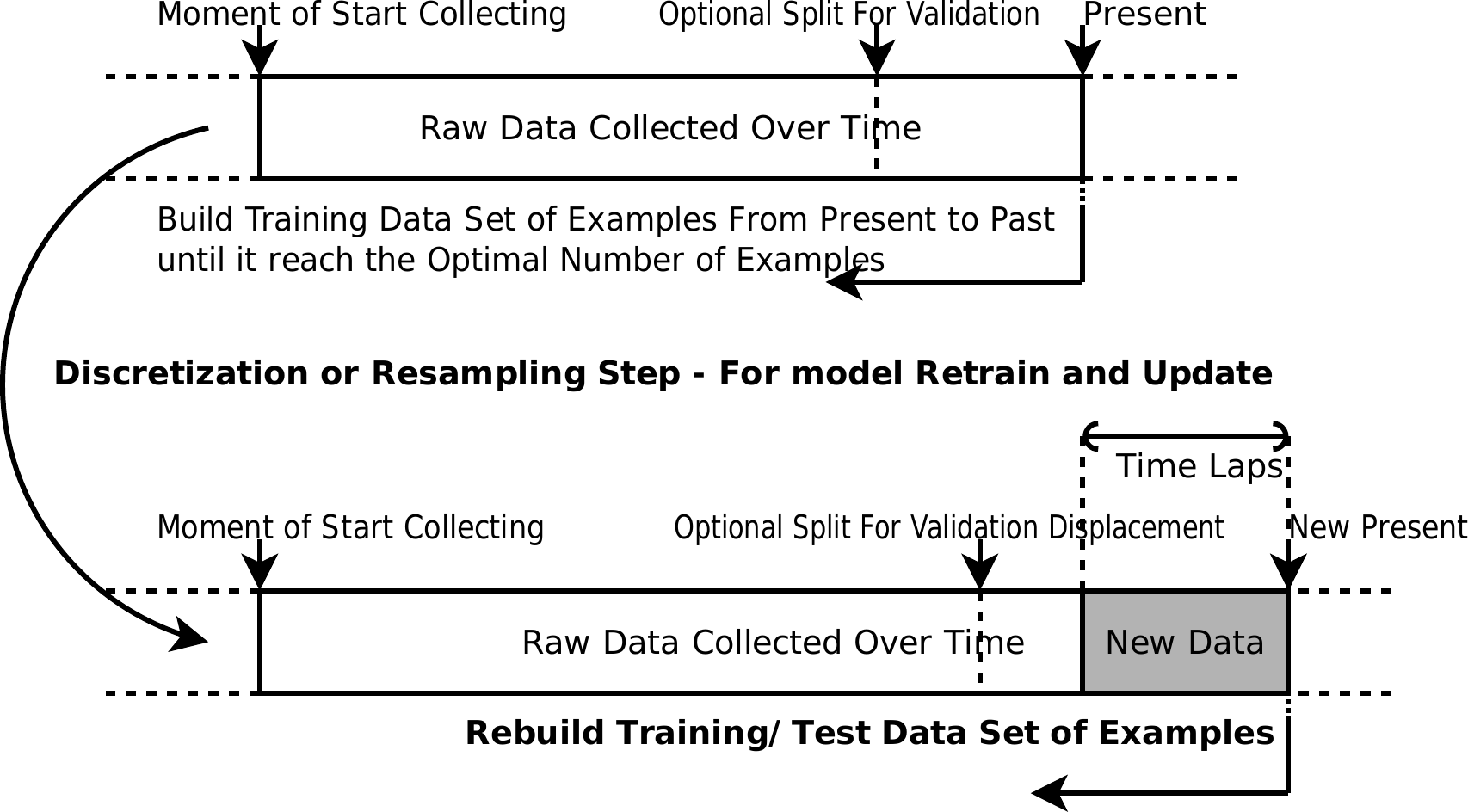}
		\caption{Add-in information to the raw dataset for global re-training of the models.}
		\label{fig-framework}
	\end{center}
\end{figure}

Raw data were collected twice per second. There are between 15 to 25 races per day, and data are collected only during the 10 minutes before a race starts. The goal is to process the collected data and update the models every month, as shown in Fig. \ref{fig-framework}. The raw data correspond to the data frames listed in Table \ref{table-ladder}, which were collected and then transformed into examples used to fit the models. These are organized into training and test datasets. The set of examples is constructed by going from the present to the past until the maximum number of examples defined for the training purpose is reached.

\subsubsection{Rule-Base Filtering}\label{cat}
This step required expert opinions on the specifics of this type of market \citep{GONCALVES201938}. Table \ref{table-rule} systematizes the decision tree that underlies the dynamics of each market. The combination of all properties generates 54 distinct categories (i.e., tree leaves), which are indexed to facilitate data processing. For instance, category 41 corresponds to a market dynamic characterized by the following properties:
 $$ 
 {\rm
root/nofavorite/mediumRunners/midleOdd/highLiquidity/ \Rightarrow Model(41)
}
$$
Out of the 54 categories, only 9 satisfy the minimum data requirements necessary to train the models, as these correspond to the more likely market states. To train a model from scratch, we define the minimum number of examples required as 1200. For the remaining categories, further studies must be conducted to explore the use of transfer learning. Transfer learning should be performed sequentially, moving from one category to the next based on similarity.

\begin{table}[htbp]
	\begin{center}
		
		\small
		\begin{tabular}{|c|c|c|c|}	
			\hline
			\textbf{Favorite} & \textbf{Runners} &\textbf{Price }& \textbf{Liquidity}\\
			\textbf{(1)} & \textbf{(2)} &\textbf{(3)}& \textbf{(4)}\\
			\hline
			\hline
			Yes & Few &  High & High \\
			\hline
			No & Medium &  Medium & Medium \\
			\hline
			\multicolumn{1}{|c|}{\textbf{}} 	  & Many & Low & Low \\
			\hline
		\end{tabular}
        \includegraphics[width=0.7\textwidth]{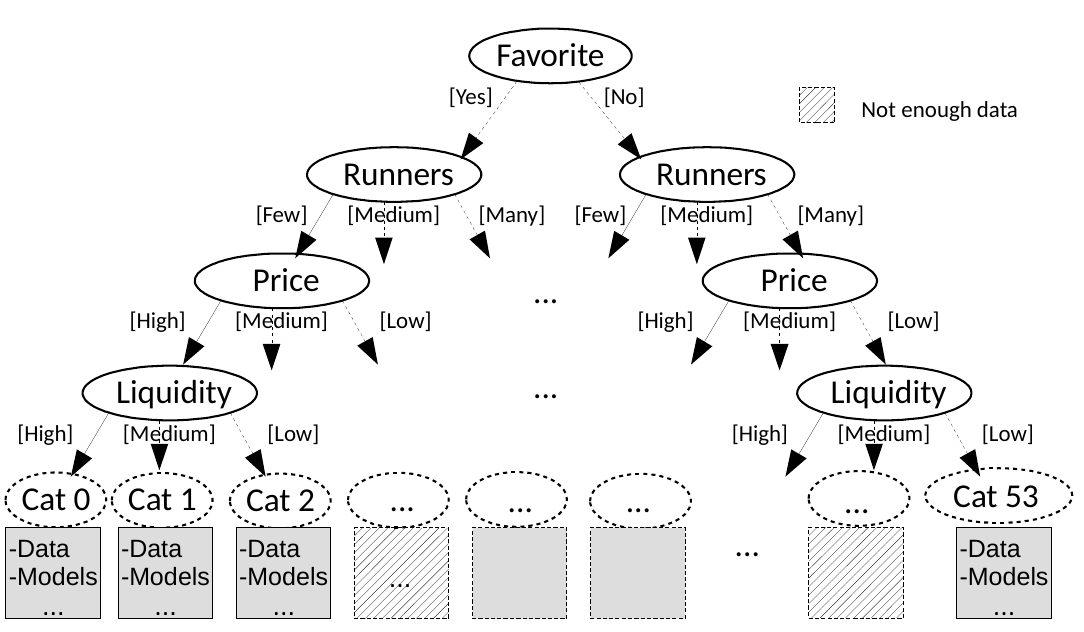}
		\caption{Rule-based decision tree. }
		\label{table-rule}
	\end{center}
\end{table}

\subsubsection{Input and Output Variables}

\begin{figure}[H]
	\begin{center}
		\leavevmode
		\includegraphics[width=0.9\textwidth]{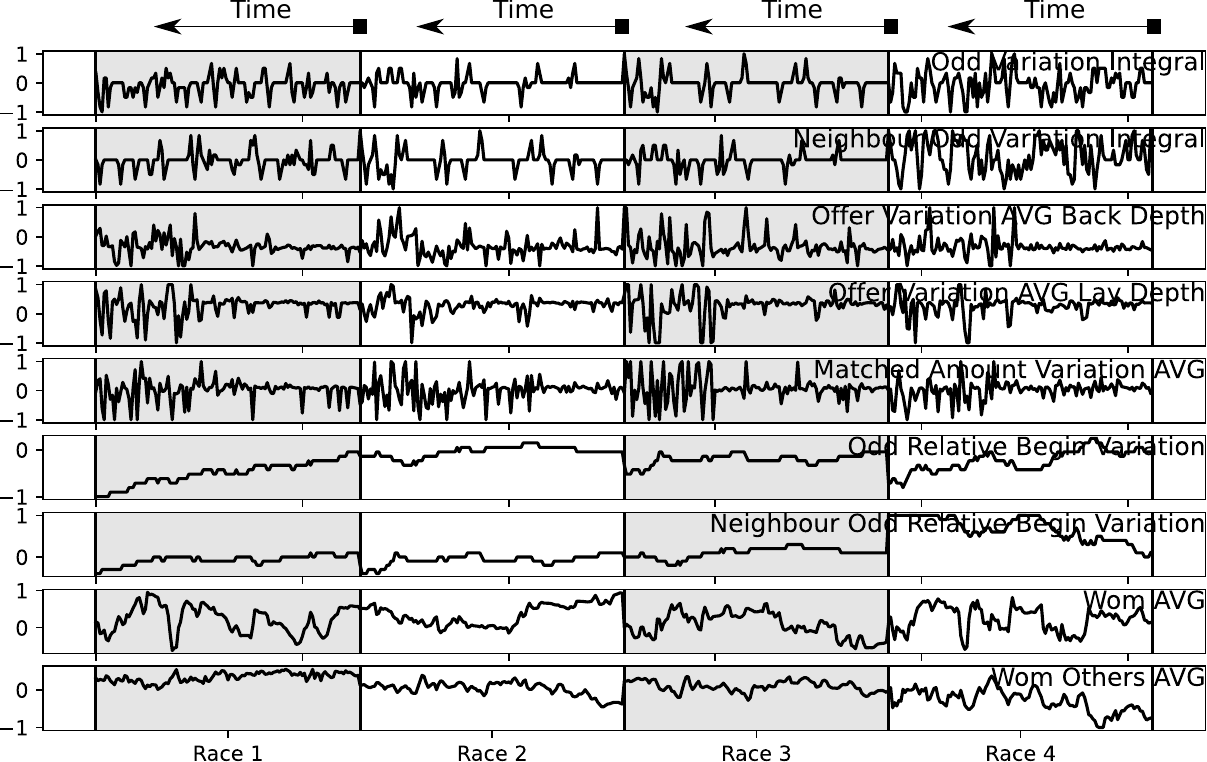}
		\caption{4 races input examples, indicators evolution. Each input sample has 4.5 minutes of data.}
		\label{fig-example-races-indicators}
	\end{center}
\end{figure}

\begin{figure}[H]
	\begin{center}
		\addtolength{\tabcolsep}{-4pt}
		\begin{tabular}{ccccc}
			\includegraphics[width=0.90\textwidth]{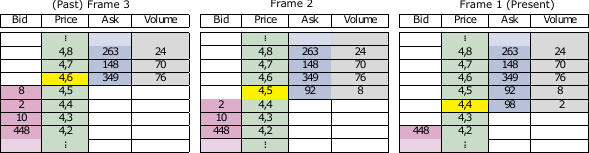}
			\\
			\tiny
			\begin{minipage}{.22\linewidth}
				\center{Indicator 1 (and 2)}
				\begin{tabular}{|c|c|c|}
					\hline
					Frame 3&Frame 2&Frame1\\
					\hline
					\hline
					4,6 &&\\
					\hline
					&\cellcolor{yellow}4,5&\cellcolor{yellow}\\
					\hline
					& &\cellcolor{yellow}4,4 \\
					\hline
				\end{tabular}
				\center{\center{$\int=-3$}}
			\end{minipage}%
			\begin{minipage}{.22\linewidth}
				\center{\ \ \ Indicator 3 \ \ \ }
				\begin{tabular}{|c|c|c|}
					\hline
					Frame 3&Frame 2&Frame1\\
					\hline
					\hline
					\cellcolor{myblue}0 & &\\
					\hline
					&\cellcolor{myblue}+92&\\
					\hline
					& &\cellcolor{myblue}+98 \\
					\hline
				\end{tabular}
				\center{\center{$\sum=+190$}}
			\end{minipage} 
			\begin{minipage}{.22\linewidth}
				\center{Indicator 4}
				\begin{tabular}{|c|c|c|}
					\hline
					Frame 3&Frame 2&Frame1\\
					\hline
					\hline
					\cellcolor{mypink}0 & &\\
					\hline
					&\cellcolor{mypink}-8&\\
					\hline
					& &\cellcolor{mypink}-2-10 = -12 \\
					\hline
				\end{tabular}
				\center{\center{$\sum=-20$}}
			\end{minipage} 
			\\
			
			\\
			\tiny
			\begin{minipage}{.22\linewidth}
				\center{\ \ \ Indicator 5 \ \ \ }
				\begin{tabular}{|c|c|c|}
					\hline
					Frame 3&Frame 2&Frame1\\
					\hline
					\hline
					\cellcolor{gray!30}0 & &\\
					\hline
					&\cellcolor{gray!30}-8&\\
					\hline
					& &\cellcolor{gray!30}-2 \\
					\hline
				\end{tabular}
				\center{\center{$\sum=-10$}}
			\end{minipage} 
			\begin{minipage}{.22\linewidth}
				\center{Indicator 6 (and 7)}
				\begin{tabular}{|c|c|c|}
					\hline
					Frame 3&Frame 2&Frame1\\
					\hline
					\hline
					\cellcolor{green!10}4.6 & &\\
					\hline
					&\cellcolor{yellow}4.5&\\
					\hline
					& &\cellcolor{yellow}4.4 \\
					\hline
				\end{tabular}
				\center{\center{$Diff\text{ }Ticks=-2$}}
			\end{minipage} 
			\begin{minipage}{.19\linewidth}
				\center{Indicator 8 (and 9)}
				\begin{tabular}{|c|c|c|}
					\hline
					Frame 3&Frame 2&Frame1\\
					\hline
					\hline
					\cellcolor{orange!50}0.02 & &\\
					\hline
					&\cellcolor{orange!50}0.43&\\
					\hline
					& &\cellcolor{orange!50}.45 \\
					\hline
				\end{tabular}
				\center{\center{$AVG(\gls{WoM})=0.43$}}
			\end{minipage} 
		\end{tabular}
	\end{center}
	\caption[Caption for LOF]%
	{Example of processing the 9 indicators given a segment of 3 \gls{RDF}.}
	\label{fig-indic-1}
	\caption*{Note: In frame 1 of Fig. \ref{fig-indic-1}, the Lay amount of 10 at 4.3 disappears not due to the matching process, but rather to exemplify a cancellation of the amount.}
\end{figure}

For the input set, 512 \glspl{RDF} are used, which corresponds to approximately 4.5 minutes of data to predict the price movement in the last 2 minutes before the start of a race. This means we need at least 6.5 minutes of pre-live market data for each race. Segments of 4 \gls{RDF} are compressed into a single value to build indicators (cf. Fig. \ref{fig-indic-1}), resulting in 128 time steps. This data compression helps with computation and reduces the risk of overfitting. Each race constitutes one example for training. Nine indicators are used as model inputs, leading to the input format: \textit{128 TimeSteps $\times$ 9 Variables}. Fig. \ref{fig-example-races-indicators} shows 4 examples (i.e., 4 races with the evolution of the 9 indicators). The 9 indicators selected as input variables for the \gls{DL} \gls{NN} models are:
\begin{enumerate}
	\item Integral of the price change of the runner in trade;
	\item Integral of the price change of the competitor runner;
	\item Liquidity variation on the ask side;
	\item Liquidity variation on the bid side;
	\item Volume variation and direction;
	\item Price variation relative to the beginning of the sequence of the runner in trade;
	\item Price variation relative to the beginning of the sequence of the competitor runner;
	\item \gls{WoM} of the runner in trade; and
	\item \gls{WoM} of all other runners combined.
\end{enumerate}

\noindent Fig. \ref{fig-indic-1} exemplifies a market state evolution corresponding to $3$ \glspl{RDF} in discrete time periods, compressed into one time segment value. It serves as a showcase for constructing the indicators considered in this case study. Selected indicators are also presented in Fig. \ref{fig-indic-1}. The first indicator corresponds to the integral of the price change of the runner subject to modeling, given by the integration of the ticks during the time segment. The second indicator corresponds to the integral of the price change of the competitor runner. In our case, the competitor runner is the one holding the closest price. In financial markets, this choice may be determined by expert opinion, i.e., another market with a strong positive or negative correlation. Since this is similar to the previous indicator, its graphical representation is omitted. Third and fourth indicators correspond to the variation in the amount on the Ask and Bid sides, respectively. Note that, for this example, it is assumed that the fourth past frame is equal to the third past frame. The fifth indicator highlights the \textit{market strength}. It is given by the variation of the matched amounts, which provides information about the volume direction and strength. The sixth indicator corresponds to the price variation between the beginning of the entire sequence $t_0$ and the segment in processing $t_i$. For the numerical example exposed in Fig. \ref{fig-indic-1}, we assume that this $3^\text{rd}$ \gls{RDF} segment is the first of the 4.5-minute range. The seventh indicator is the same but applied to the competitor runner. The eighth indicator is the average \gls{WoM} of the \glspl{RDF} in the segment. The \gls{WoM} is represented by a percentage value and shows when the market is balanced or unbalanced. The market is said to be balanced when the amount of money unmatched on each side of a selection is the same. This means the amount placed on the ask side must be approximately equal to that placed on the bid side.

 The underlying logic is straightforward: when there is more unmatched money on the ask side than the bid side, the price decreases. The \gls{WoM} pushes the price down. The same applies the other way around. The \gls{WoM} indicator is given by:
\begin{align}
\label{eq:wom}
{\rm 
WoM = \frac{Amounts\text{ }Bid} {Amounts\text{ }Bid - Amounts\text{ }Ask 
}
}
\end{align}
For the numerical example shown in Fig. \ref{fig-indic-1}, we consider only the depth of the best 3 prices around the transacted price. Depending on whether the price is high, medium, or low (Table \ref{table-rule}), the depth used is 2, 3, and 4, respectively. For the ninth indicator, \gls{WoM} is also applied but to all other runners. The logic is that if the ladder of unmatched amounts in all other runners is pressing the price in one direction, the runner in trade will be pressed in the opposite direction. Finally, the model output or \textit{target} corresponds to the integral of the price variation, measured in ticks, for the last two minutes before the race starts. By compressing the data from multiple segments in this manner, we define a \gls{MTS} problem with 128 time steps.

\subsubsection{Frequency Distribution Histograms}
\begin{figure}[H]
	\centering
	\begin{subfigure}{1\textwidth}
		\centering
		\includegraphics[width=0.8\textwidth]{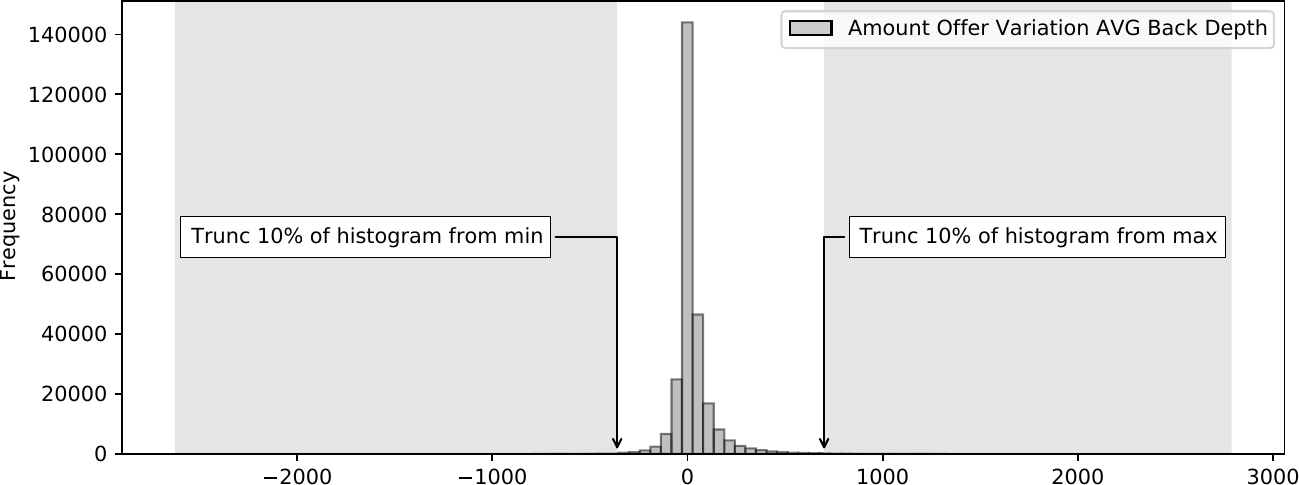}
		\label{fig:up}
	\end{subfigure}
	\begin{subfigure}{1\textwidth}
		\centering
		\includegraphics[width =0.8 \textwidth]{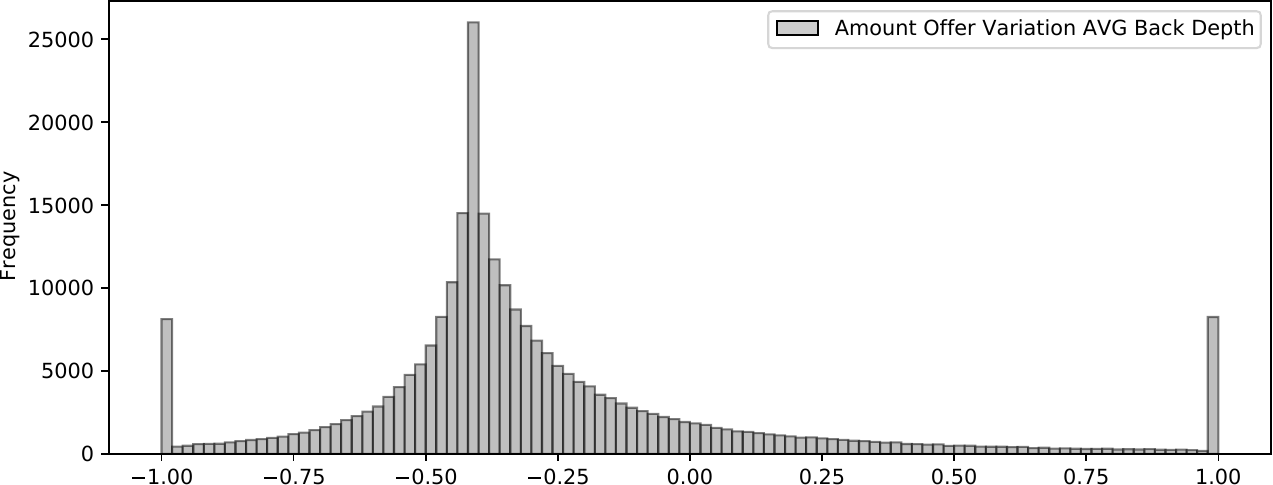}
		\label{fig:down}
	\end{subfigure}
	\caption[Histogram re-scaling with truncated tails at $10\%$ level to find min-max values for input normalization.]{Example of histogram re-scaling with truncated tails at $10\%$ level to find min-max values for input normalization. The illustration represents input $\#3$ (Liquidity variation on the ask side) on category $\#41$ of the rule-based system index.}
	\label{fig-hist-rescale}
\end{figure}

To address outlier detection, we apply the truncation technique described in \cite{deboeck1994trading}. Fig. \ref{fig-hist-rescale} illustrates the automatic process of outlier truncation. Afterwards, the data are normalized into the interval $[-1,1]$ through frequency analysis and histogram rescaling. The rescaling of maximum and minimum raw values involves truncating 10\% of the histogram tails. This operation alters, but does not remove, the original examples. Only after these steps are the data fed as input for model training. This operation is systematically applied to all inputs for each category.\footnote{Fig. \ref{fig-hist-horses-apx} in Appendix shows the result of this operation for all indicators used in this case study.}

\begin{figure}[htbp]
	\begin{center}
		\leavevmode
		\includegraphics[width=0.7\textwidth]{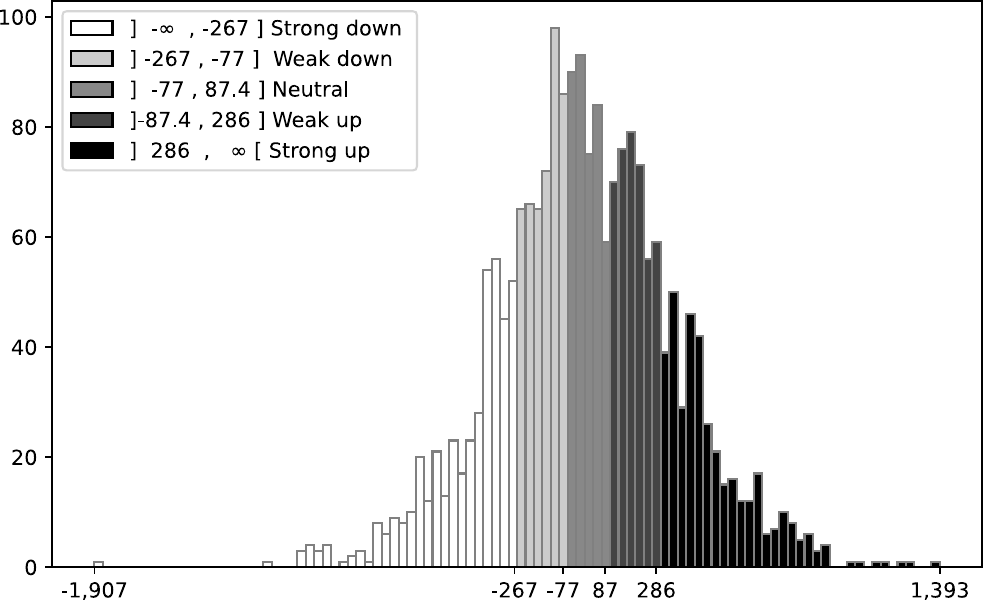}
		\caption[Histogram for the qualitative classification of the output representing the integral of tick variation.]{Example of the histogram for the qualitative classification of the output representing the integral of tick variation for the last 2 minutes: strong down (left white), weak down (light gray), neutral (gray), weak up (dark gray), and strong up (right black), respectively.}
		\label{fig-hist-output}
	\end{center}
\end{figure}

A similar technique, using histograms for data partitioning, is applied at the output level to transform the regression problem into a classification problem. An egalitarian distribution of examples across qualitative classes is ensured to avoid overfitting and/or biased models. The output for this case study is the integral (i.e., the area) of the tick variation of the runner's price relative to the last 2 minutes before the race starts. When the integral is significantly negative, a ``strong down'' price change is considered, and the first qualitative class is established. If the numerical value falls into the second qualitative class, a ``weak down'' price change is assumed. In the third qualitative class, a ``neutral'' price change is considered. A ``weak up'' price change is assigned when the solution falls into the fourth qualitative class. Finally, a ``strong up'' price change occurs when the solution falls into the fifth qualitative class. This classification procedure is outlined in Fig. \ref{fig-hist-output}. Each class will contain approximately 20\% of the examples in the training dataset. A ``strong'' movement prediction suggests activating a trailing-stop trading mechanism, while a ``weak'' movement prediction indicates the activation of a small swing trading strategy.

 The choice of target and stop-loss prices in each category is based on the de-normalization of the output, as shown in the histogram presented in Fig. \ref{fig-hist-output}. The main idea behind this process is to adjust the parametrization of trading mechanisms -- target and stop-loss -- according to each category. The target price is determined as the average of the maximum tick variation for all examples within a given class during the predicting time. The stop-loss is then defined as 80\% of the target price for swing trades and 60\% of the target price for trailing-stop trades. Table \ref{table-param-trading} exemplifies how target and stop-loss prices are defined for each class within a given category.

\vspace{.3cm}

\begin{table}[htbp]
	\begin{center}
		\small
		\begin{tabular}[t]{|c|c|c|c|}
			\hline
			Class & Mean of the ticks variation & Target & Stop-loss \\ \hline 
			Strong Up & 6.44794 & 6 & 4 \\ \hline
			Weak Up & 3.51428 & 4 & 3\\ \hline 
			Weak Down & -3.19424 & 3 & 2\\ \hline
			Strong Down & -6.33173 & 6 & 4\\ \hline		
		\end{tabular}
	\end{center}
	\caption{Example of trading mechanism parameters for a particular category.}
	\label{table-param-trading}
\end{table}

\noindent Once the developed \gls{DL} \gls{NN} models are ready for production, they undergo final validation in the simulator described in Subsection \ref{sec-software-framework}. To provide a better understanding of the trading execution based on the model's predictions, Table \ref{table-trading} presents log results of a single trading execution.

\begin{table}[htbp]
	\begin{center}
		
		\footnotesize
		\begin{tabular}[t]{|c|c|}
			\hline
			\multicolumn{2}{|c|}{Instantiation}\\
			\hline 
			Runner category & 
			$nofavorite/mediumRunners/midleOdd/highLiquidit$ (Model \#41)  \\
			\hline 
			Model predicted probabilities  & [0.14 , 0.19, 0.17 , 0.20 , {\bf0.30}] \\ \hline 
			Predicted class  & $5^{th}$ class: Strong Up \\ \hline
			Bets/trade direction & Up: Lay (open) $\Rightarrow$ Back (close)\\ \hline
			Trading mechanism & Trailing Stop (strong movement predicted) \\ \hline 
			Parameters (in ticks) & Stop-loss: 4 , Target: 6 \\ \hline
			Parameters (in odds) & Entry odd: 4.6 , Target odd: 5.2 , Stop odd: 4.2 \\
			\hline		
			Time parameters & 20 frames open, 80 frames start close best price, 20 close emergency  \\ \hline
			Open amount stake & $\pounds{3.00}$ (Lay) \\ \hline
			Potential \gls{PL} &  Profit: $\pounds{0.35}$ , Loss: $-\pounds{0.28}$  \\ \hline
			\hline			
			\multicolumn{2}{|c|}{Result}\\
			\hline
			Trade final state & CLOSED \\ \hline
			Moved ticks  & 6 \\ \hline
			Open amount stake & $\pounds{3.00}$ (Lay) \\ \hline
			Effective open odd (price) & 4.6 \\ \hline
			Close amount stake & $\pounds{2.65}$ (Back) \\ \hline
			Effective \gls{PL}  & $\pounds{0.35}$ \\ \hline
			Effective close odd (price) & 5.2 \\ \hline

		\end{tabular}
		\caption{Information example of one trading execution log line with the model prediction, category parameters, and results. Table \ref{table-horse-final-output} in the Appendix shows a segment of the output logs obtained.}
		\label{table-trading}
	\end{center}
\end{table}


\section{Methods: Applied Deep Learning Architectures}\label{sec-3}
Let us now explain in detail the \gls{DL} architectures that form the basis of this study, along with the proposed extensions.

\subsection{CNN LeNet Based Models} \label{sec-cnn-basic-lenet}
\begin{figure}[htbp]
	\begin{center}
		\leavevmode
		\includegraphics[width=0.34\textwidth]{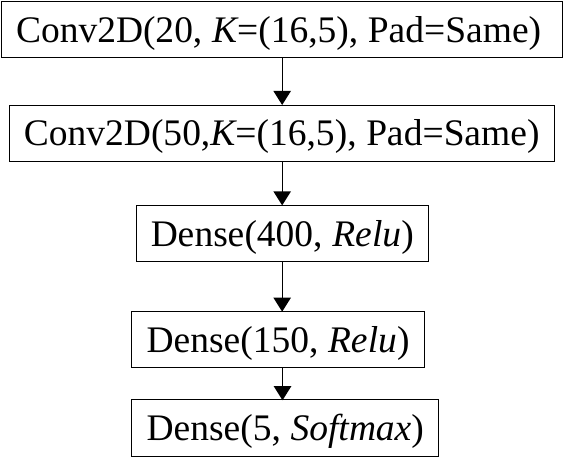}
		\caption{Base LeNet \gls{CNN} 2D using same padding in convolutional layers}
		\label{fig-LeNet}
	\end{center}
\end{figure}

\cite{726791} proposed a \gls{NN} architecture for handwritten and machine-printed character recognition, called LeNet. This architecture, based on convolutional layers, is a straightforward \gls{CNN} that is easy to understand. The LeNet-5 architecture consists of two sets of convolutional and average pooling layers, followed by a flattening operation and three dense layers (cf. Fig. \ref{fig-LeNet}). \glspl{CNN} are also described in detail in \cite{5537907}. This simple model is introduced here as a formalization and should be interpreted as a basic benchmark case relative to alternative architectures. It is also used to identify improvements with the add-ons further described.

Consider a bi-dimensional input feature map \(x^l\) in layer \(l\) of size \((H, W)\) and a stride \(\delta\) of \((1, 1)\). The basic mathematics for the computation of a convolutional layer \(l\) to obtain the output feature map \(y^l\) with kernel \(K\) of size \((k_H, k_W)\) can be expressed as:
\begin{align}\label{eq:cnn-compute}
	{y}^{l}_{i,j} = \phi^l \left( \sum_{i=0}^{H-{k_H}} \sum_{j=0}^{W-{k_W}} K \cdot {x^l}_{i,j} \right)
\end{align}
where \(\phi\) is the activation function. This equation represents the application of kernel \(K\) in the input map \(x^l\) at coordinates \(i, j\) in layer \(l\). A bias term \(b\) is usually added to \({y}^{l}_{i,j}\), but it is omitted here for clarity. 
It is observed that the output is smaller than the input when the convolution kernel is larger than \((1, 1)\). If the input has size \((H, W)\) and the kernel \(K = (k_H, k_W)\), the resulting convolution has size \((H - k_H + 1, W - k_W + 1)\), which is smaller than the original input. Typically, this is not a concern for inputs with large dimensions (i.e., images) and small filters. However, it can be problematic with small input dimensions or when considering a high number of stacked convolutional layers. In such cases, the practical effect of large filter sizes and/or very deep \glspl{CNN} on the size of the resulting feature map could lead to loss of information, causing the model to run out of data. The padding operation is introduced to address this issue.

\subsubsection{Traditional Padding Methods}
Currently, the standard procedure to avoid the border effect problem consists of applying \textit{same padding} (i.e., the inclusion of zeros outside of the input map). For each channel of the bi-dimensional input \(x\), we insert zeros \({k_H-1}/{2}\) rows above the first row and \({k_H}/{2}\) rows below the last row, as well as \({k_W-1}/{2}\) columns to the left of the first column and \({k_W}/{2}\) columns to the right of the last column. This approach ensures that the convolution output size will be \((H, W)\), maintaining the same spatial extent as the input. 

\begin{table}[htbp]
	\begin{center}
		
		\small
		\begin{tabular}{|c|c|c|c|}	
			\hline
			&\multicolumn{3} {c|}{Example of padded info}   \\
			\hline
			Method & Pad& Input & Pad \\
			\hline
			Valid (None) & &\fontfamily{qcr}\selectfont a b c d e f & \\
			\hline
			Same (Zero) &\fontfamily{qcr}\selectfont 0 0 0 0 &\fontfamily{qcr}\selectfont a b c d e f &\fontfamily{qcr}\selectfont 0 0 0 0\\
			\hline
			Reflect (Mirror) & \fontfamily{qcr}\selectfont d c b a &\fontfamily{qcr}\selectfont a b c d e f &\fontfamily{qcr}\selectfont f e d c\\
			\hline
			Reflect101 &\fontfamily{qcr}\selectfont e d c b &\fontfamily{qcr}\selectfont a b c d e f &\fontfamily{qcr}\selectfont e d c b\\
			\hline
			Constant n &\fontfamily{qcr}\selectfont n n n n &\fontfamily{qcr}\selectfont a b c d e f &\fontfamily{qcr}\selectfont n n n n\\
			\hline
			Tile 2 &\fontfamily{qcr}\selectfont a b a b &\fontfamily{qcr}\selectfont a b c d e f &\fontfamily{qcr}\selectfont e f e f\\
			\hline		
			Causal (Zero Left) &\fontfamily{qcr}\selectfont 0 0 0 0 &\fontfamily{qcr}\selectfont a b c d e f &\\
			\hline		
			Wrap &\fontfamily{qcr}\selectfont c d e f &\fontfamily{qcr}\selectfont a b c d e f &\fontfamily{qcr}\selectfont a b c d\\
			\hline		
			
		\end{tabular}
		\caption{Padding examples of size 4 for unidimensional input. }
		\label{table-pads}
	\end{center}
\end{table}

 However, when analyzing a \gls{MTS} problem, where the input feature map has a relatively small size in the variable component, the inclusion of zeros via same padding may weaken the learning capability. This is because the learned kernel interacts with the zero values in the input during the dot product operation, potentially leading to erroneous generalization and less effective model training. According to \cite{7371214}, there are other well-known padding methods commonly used in image processing environments. These use the information in input variables to fill in the borders. Table \ref{table-pads} provides examples of different padding methods.

\subsubsection{A New Method: Roll Padding}\label{sec-roll}
Roll padding is an extension of wrap padding, specifically designed for \gls{MTS} analysis. Wrap padding copies information from the opposite sides of the image, effectively mapping it onto a torus. This operation results in four copies of information under a bi-dimensional input: rows (columns) above the top (on the left) are duplicated from the bottom rows (right columns), respectively, and vice versa. While wrap padding is generally not useful for natural images, it is highly effective for computed images, such as Fourier transforms and polar coordinate transforms, where pixels on opposite borders are computationally adjacent.

\begin{figure}[htbp]
	\begin{center}

		\begin{minipage}{0.45\linewidth}
			\includegraphics[width=0.95\textwidth]{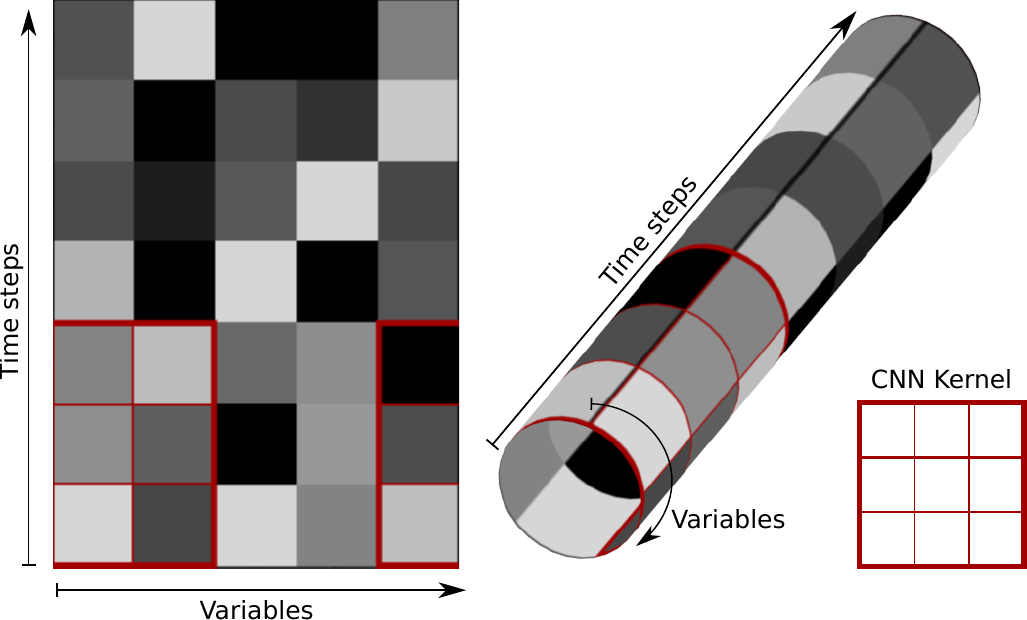}
			\hfill
		\end{minipage}
		\hspace{0.01\linewidth}
		\begin{minipage}{0.30\linewidth}
			\footnotesize
			\begin{tabular}{|c|c|c|c|}
				\hline
				& Roll & Variables & Roll \\
				\hline                                  
				\rotatebox[origin=l]{90}{Valid } &  &  &  \\
				\hline                                  
				&\fontfamily{qcr}\selectfont cdef&\fontfamily{qcr}\selectfont abcdef& \fontfamily{qcr}\selectfont abcd \\
				&\fontfamily{qcr}\selectfont ijkl&\fontfamily{qcr}\selectfont ghijkl&\fontfamily{qcr}\selectfont ghij\\
				\multirow{-3}{*}{\rotatebox[origin=l]{90}{Time }} &\fontfamily{qcr}\selectfont opqr&\fontfamily{qcr}\selectfont mnopqr&\fontfamily{qcr}\selectfont mnop\\
				\hline                                  
				\rotatebox[origin=l]{90}{Valid }& & &  \\
				\hline
			\end{tabular}
		\end{minipage}
	\end{center}
	\caption{Roll padding scheme in \gls{MTS} analysis.}
	\label{fig-roll-pad}
\end{figure}

As illustrated in Fig. \ref{fig-roll-pad}, roll padding extends this concept by copying information from the opposite sides but only along a single dimension -- input variables component in the bi-dimensional input map (i.e., \textit{TimeSteps $\times$ Variables}). In contrast, the time steps component remains unpadded (i.e., valid), transforming the structure into a cylinder rather than a torus. The reduction of the time steps component after multiple convolutions is generally not problematic, given that \gls{MTS} problems often involve a high number of time steps. However, for cases where preserving temporal resolution is crucial, roll padding can be combined with other padding methods (e.g., causal padding), as demonstrated in Fig.~\ref{fig-roll-causal-pad} and Subsection \ref{sec-WaveNet}.

\begin{figure}[htbp]
	\begin{center}
		\leavevmode
		\includegraphics[width=0.4\textwidth]{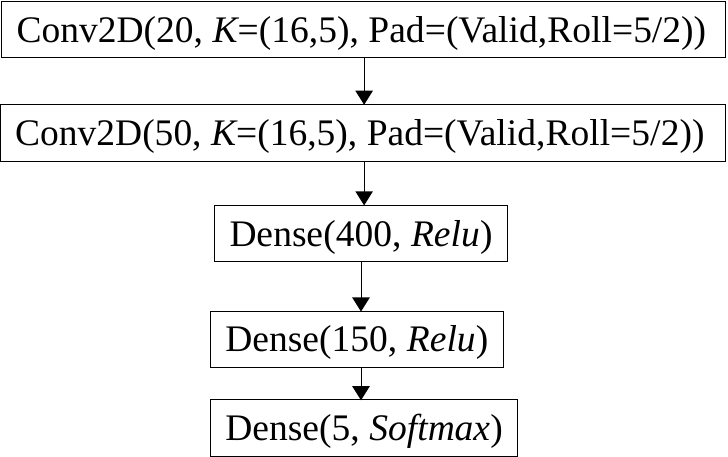}
		\caption{\gls{CNN} 2D using valid padding in time steps component and roll pading, of size $\frac{K_H}{2}$, in the variables component.}
		\label{fig-LeNet-roll}
	\end{center}
\end{figure}

As shown in Fig. \ref{fig-LeNet-roll}, the structure of \glspl{CNN} using roll padding is based on LeNet-5 \citep{726791}. Both \glspl{CNN} share the same hyperparameters, with the only difference being the padding method applied. This allows us to directly assess the impact of roll padding on performance. In Fig. \ref{fig-LeNet-roll}, we specify the padding method used in each dimension for the second \gls{CNN}. The time steps component employs valid padding, while roll padding is applied to the variables component.

\subsection{LSTM Based Models}
\begin{figure}[htbp]
	\begin{center}
		\leavevmode
		\includegraphics[width=0.26\textwidth]{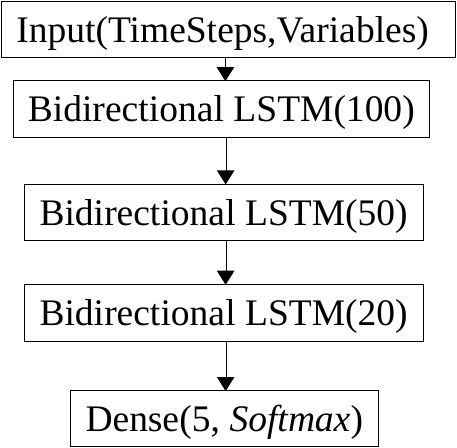}
		\caption{Stacked Bidirectional \gls{LSTM}s. LSTMs based models baseline}
		\label{fig-lstm}
	\end{center}
\end{figure}

In this study, the \gls{LSTM}, originally proposed by \cite{hochreiter1997long}, consists of four layers, as illustrated in Fig. \ref{fig-lstm}. Three of these layers use bidirectional \gls{LSTM}s, while the final layer is a dense layer with a softmax activation function for classification. Similar to the model discussed in Subsection \ref{sec-cnn-basic-lenet}, this architecture serves as a benchmark to assess improvements when integrating additional methodologies. \glspl{RNN} have been successfully applied to numerous sequential data tasks. Enhanced models, such as \gls{LSTM}s, facilitate training on long sequences by addressing issues like vanishing gradients. However, despite these advancements, even the most sophisticated models face limitations, making it challenging for researchers to develop high-quality solutions for long-sequence data. Many \gls{MTS} problems require establishing connections between distant input and output points across multiple layers, often spanning dozens of time steps. To effectively tackle such challenges, existing \gls{RNN} architectures have had to be modified and adapted.

\subsubsection{Standard Attention}
Attention is a mechanism designed to be integrated with \gls{RNN}s, allowing the model to focus on specific parts of the input sequence when predicting certain parts of the output sequence. This capability enhances both the speed and robustness of convergence. The incorporation of attention mechanisms has significantly improved performance across various tasks, making it an essential component of modern \gls{RNN} networks.  

A pivotal study on attention is \cite{vaswani2017attention}, which originally introduced the mechanism for machine translation tasks. However, attention has since been widely adopted across numerous application domains. Fundamentally, attention can be viewed as a residual block that multiplies its output with its own input, \( h_i \), before reconnecting to the main \gls{NN} pipeline through a weighted and scaled sequence. Scaling parameters, known as attention weights \( \alpha_i \), determine the importance of each input, while the resulting weighted sum is referred to as the context weight \( c_i \) for each sequence position \( i \). Collectively, these form the context vector \( c \), which has a sequence length of \( n \). This operation is mathematically expressed as follows:  
\begin{align}\label{eq:ateention}
c_i = \sum_{i=0}^{n} \alpha_i  h_i
\end{align}

The computation of $\alpha_i$ results from applying a softmax activation function to the input sequence $x^l$ on layer $l$:
\begin{align}\label{eq:act-softmax-attention}
\alpha_i = \frac{\exp(x^l_i)}{\sum\limits_{k=1}^n \exp(x^l_k)}
\end{align}
This means that input values of the sequence compete with each other for attention. Since attention scores are obtained through a softmax activation function, their sum is always 1, ensuring that scaling values in the attention vector \( \alpha \) fall within the range \([0,1]\). The mechanism described above is known as \textit{soft attention} because it is a fully differentiable and deterministic process that integrates seamlessly into a backpropagation-based system. In this approach, gradients propagate through the attention block just as they do through the rest of the network.  

Differently, \textit{hard attention} does not use a weighted average. Instead, it treats \( \alpha_i \) as a sampling probability that determines whether \( x_i \) is included in the context vector. This method replaces deterministic computation with stochastic sampling. To correctly compute gradient descent during backpropagation, \textit{hard attention} employs the Monte Carlo method, where multiple sampling iterations are performed, and their results are averaged \citep{xu2015show}. The accuracy of this approach depends on the number and quality of the samples.  

On the other hand, \textit{soft attention} follows a simpler, more conventional backpropagation approach when computing gradients within the attention block. However, its accuracy depends on the assumption that a weighted average is a good representation of the relevant input areas. Both methods have their strengths and weaknesses. Currently, \textit{soft attention} is more widely used due to its seamless integration with backpropagation, making it more efficient in practice. For this study, we exclusively use \textit{soft attention}.

\begin{figure}[htbp]
	\begin{center}
		\leavevmode
		\includegraphics[width=1\textwidth]{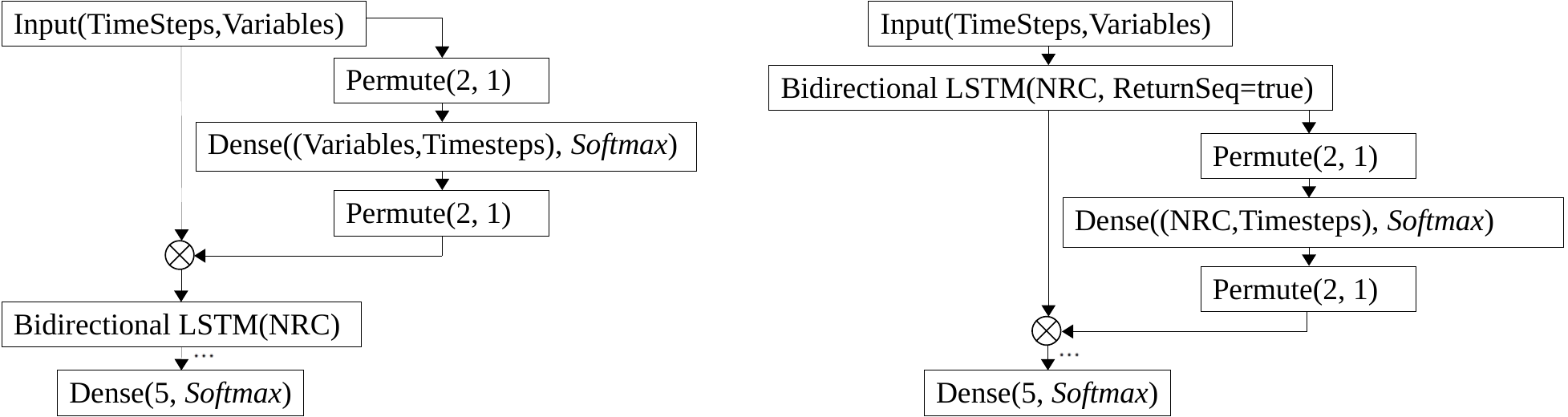}
		\caption{\gls{MTS} attention before \gls{LSTM}s on the left subplot and attention after \gls{LSTM} on the right subplot.}
		\label{fig-lstm-attention}
	\end{center}
\end{figure}

If attention is applied directly to the input before entering the \textit{attention before}. Conversely, if attention is applied to the output sequence of the \gls{LSTM}, it is called \textit{attention after}, as clarified in Fig. \ref{fig-lstm-attention}.  Since we work with \gls{MTS}, a bi-dimensional dense layer for attention is used. To ensure that the attention mechanism is applied to the time step component of each sequence rather than the variable component, we perform a permutation operation both before and after this layer. It is relevant to note that when attention is applied after the \gls{LSTM}, the recurrent layer must return its internal recursively generated sequences, which correspond to the number of units defined, denoted as \( NRC \). This parameter is crucial within the attention block, as it determines how many sequences need to be processed.

\subsubsection{Multi-Head Convolutional Attention}
A key contribution of this study is the integration of convolutional layers within the attention block. The original design of attention mechanisms was primarily intended for text processing, where attention is assigned to each embedded word individually within long sequences.  However, in \gls{MTS} problems, which are inherently more continuous and less discrete than text, it can be beneficial to focus on patterns in small contiguous segments rather than on individual values. By incorporating convolutional layers within the attention block, we enable the model to capture local temporal patterns more effectively, enhancing its ability to recognize meaningful structures within the data.

\begin{figure}[htbp]
	\begin{center}
		\leavevmode
		\includegraphics[width=1\textwidth]{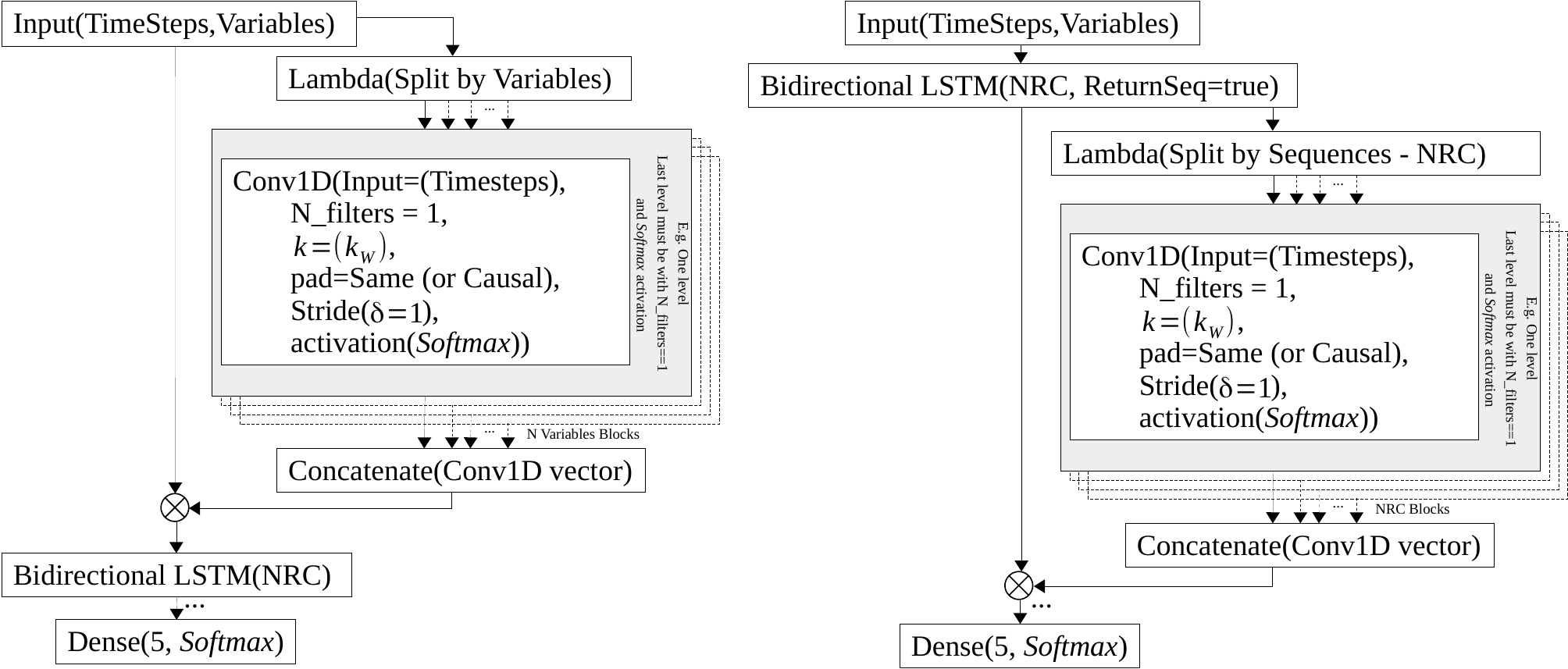}
		\caption{Attention using convolutional layers before and after \gls{LSTM}s.}
		\label{fig-cnn-attention}
	\end{center}
\end{figure}

Fig. \ref{fig-cnn-attention} shows the implementation process. The \gls{MTS} is first split into individual time series using the Keras Lambda function. For each sequence, a path with 1D convolutional layers is created, and results are concatenated back together. In Fig. \ref{fig-cnn-attention}, only one convolutional filter per sequence is depicted (i.e., per variable of the \gls{MTS}) if attention is applied before the \gls{LSTM}, or per Number of Recursive Cell (NRC) generated sequence if applied after the \gls{LSTM}. It is important to note that, before the concatenation operation, each path must return a one-dimensional vector with size \textit{TimeSteps}. When concatenated with the other paths, this results in an attention weights feature map of size \textit{TimeSteps} $\times$ \textit{Variables}. This map is then multiplied with \( h \) to obtain the 2D context map \( c \).

To capture multiple small subsequence patterns (i.e., filters), we must stack multichannel 1D convolution layers before the final attention layer. However, the last convolutional layer inside the attention block must output only a single channel, as explained earlier. An alternative way to enforce a 1D output vector for each path is to use the Keras \textit{AveragePooling1D} layer, which averages the previous channels into one dimension. Additionally, the final single-channel 1D convolution output must use the softmax activation function to ensure that each value, in the resulting vector per variable, competes with each other, sums to 1, and has a scaling factor in the \([0,1]\) range.

\subsection{ConvLSTM2D Based Models}\label{sec-ConvLSTM2D}
\subsubsection{ConvLSTM2D for Segmented Time Series} 
\begin{figure}[htbp]
	\begin{center}
		\leavevmode
		\includegraphics[width=0.65\textwidth]{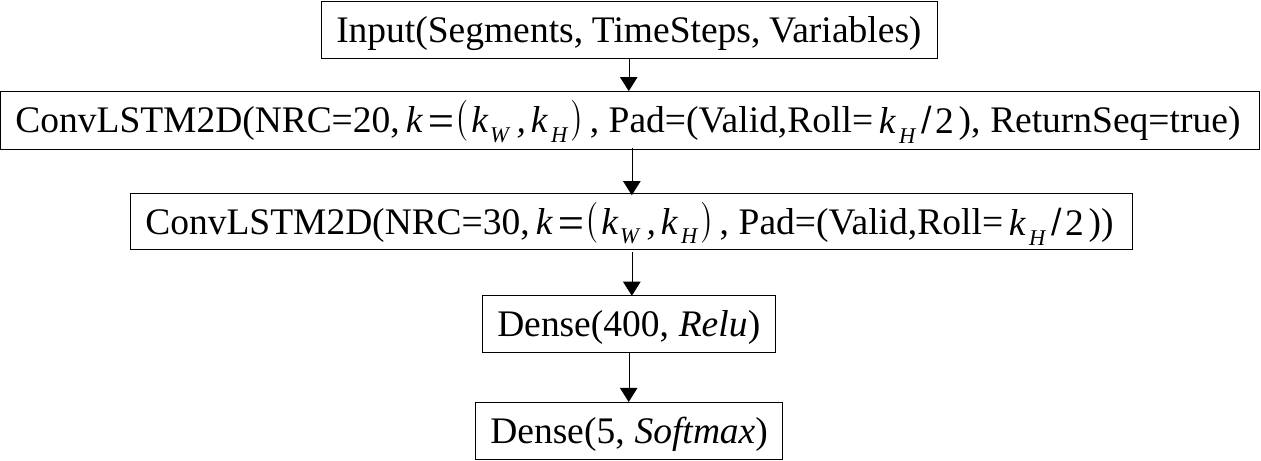}
		\caption{Base scheme for staked \gls{ConvLSTM2D} with roll padding on the variables component.}
		\label{fig-cnnlstm-stack}
	\end{center}
\end{figure}

\begin{figure}[!htbp]
	\begin{center}
		\begin{minipage}{1.0\linewidth}
			\includegraphics[width=1\textwidth]{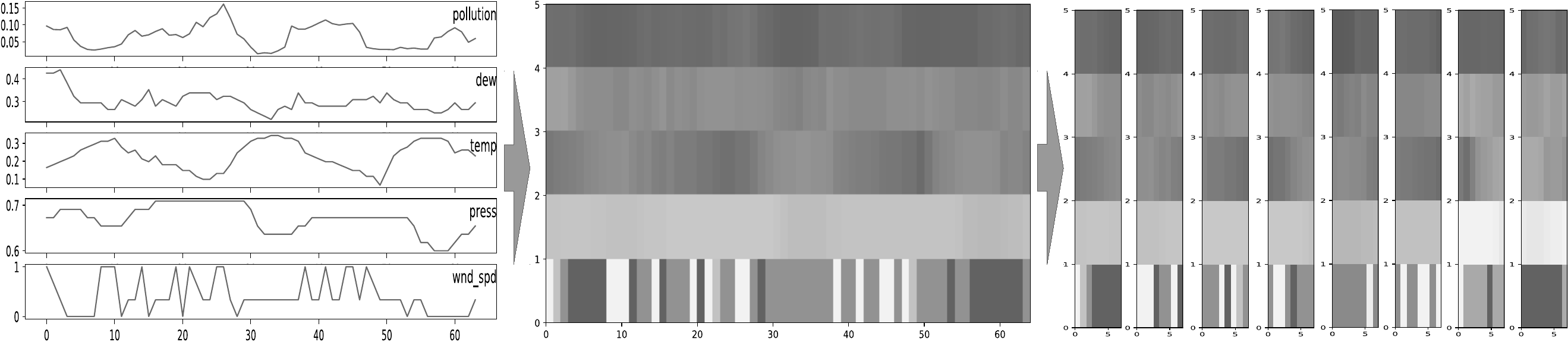}
		\end{minipage}
		\begin{minipage}{0.60\linewidth}
			\includegraphics[width=1\textwidth]{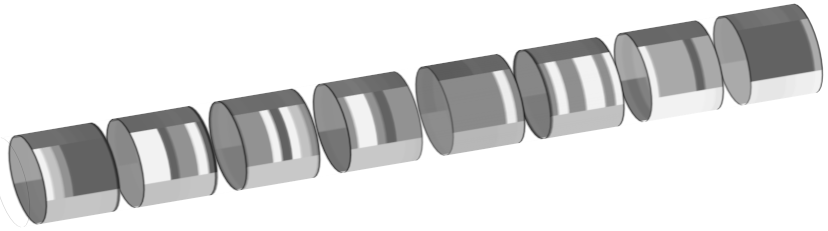}
		\end{minipage}
	\end{center}
	\caption{\gls{MTS} input processing for \gls{ConvLSTM2D}.}
	\label{fig-convlsts-input}
	\caption*{Note: The bottom plot describes the application of roll padding in the variables component for each segment.}
\end{figure}

The \gls{ConvLSTM2D} layer was proposed by \cite{shi2015convolutional}. The motivation for this structure was to predict future rainfall intensity based on sequences of meteorological images. By applying this layer in a \gls{NN} architecture, they were able to outperform state-of-the-art algorithms for this task. The \gls{ConvLSTM2D} is a recurrent layer, similar to the \gls{LSTM}, but internal matrix multiplications are replaced with convolution operations. As a result, the data flowing through the \gls{ConvLSTM2D} cells retains the input dimension, 3D in our case: \textit{Segments $\times$ TimeSteps $\times$ Variables}, rather than being just a 2D map: \textit{TimeSteps $\times$ Variables} (cf. Fig. \ref{fig-cnnlstm-stack}-\ref{fig-convlsts-input}). As explained in Subsection \ref{sec-roll}, we can also apply roll padding to this input on the variables component. \gls{ConvLSTM2D} layers can be particularly useful in \gls{MTS} that can be partitioned into segments, such as in the case study of household electric power consumption \citep{gonccalves2023variable}. In this case, the time series will exhibit representative patterns for every day of the week, which can be grouped into a 2D map.

\subsubsection{ConvLSTM2D Convolutional Attention with Roll Padding}\label{sec-att-2D-roll}

\begin{figure}[!htbp]
	\begin{center}
		\leavevmode
		\includegraphics[width=0.8\textwidth]{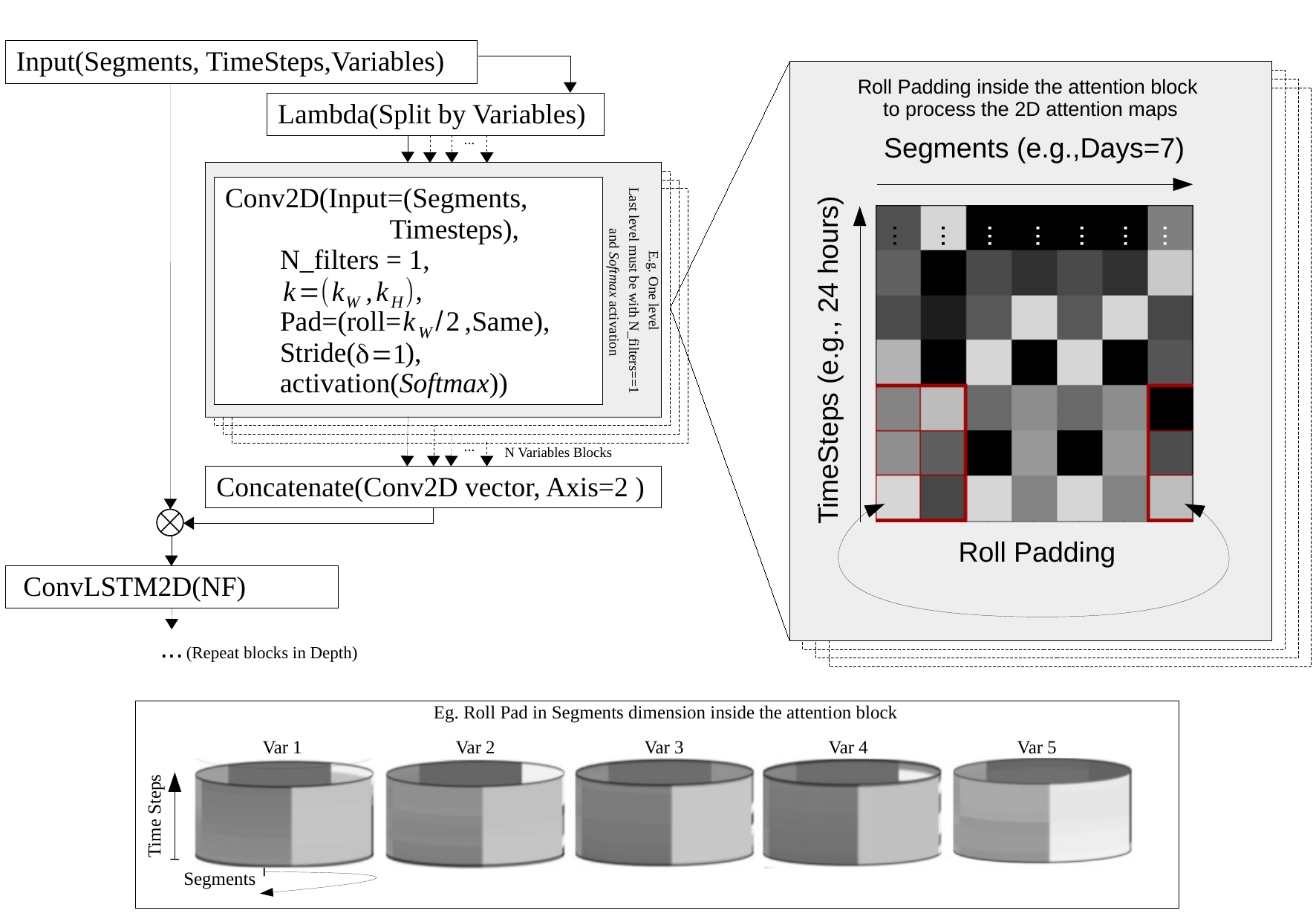}
		\caption[Multi-head Attention with 2D convolution layers before \gls{ConvLSTM2D} using roll padding in the segments component.]{Attention using 2D convolutional layers before \gls{ConvLSTM2D}.}
		\label{fig-cnnlstm-attention}
		\caption*{Note: The bottom plot describes the application of roll padding in the segments component for each variable inside the attention block.}
	\end{center}
\end{figure}

When entering the attention block, after splitting by the input variable, the resulting 2D map to be processed by convolution layers will have a \textit{Segments} $\times$ \textit{TimeSteps} format. This means that the 2D kernels will attempt to capture patterns relating to contiguous time steps, as well as the same temporal steps across the previous and next segments. If segments represent days and time steps are divided by hours, a 2D kernel will capture attention patterns related to specific hours of the day, as well as similar periods in the preceding and following days. Moreover, if we have segments of seven days, roll padding can be applied to the segments component, allowing the kernel's border processing to correlate the first day of the week with the last day, especially if the data exhibits a weekly cyclical pattern, as shown in Fig. \ref{fig-cnnlstm-attention}. If it is not desirable to correlate data between segments, a one-dimensional kernel should be defined (i.e., 2D $K = (1, k_w)$, since we are working with a bi-dimensional convolution layer). Each 2D output map is obtained through a softmax activation. Each value in the resulting 2D map for each variable competes with the others, summing to 1, with a scaling factor in the range $[0,1]$. Once all 2D maps are concatenated, the resulting $\alpha$ will be 3D and compatible for scaling the inputs $h$ of the attention block to obtain $c$, as described in Eq. (\ref{eq:ateention}).

\subsection{Multivariate WaveNet}\label{sec-WaveNet}
\subsubsection{WaveNet 1D with Multichannel Input}\label{sec-WaveNet-multi-head}
\begin{figure}[htbp]
	\begin{center}
		
		\begin{minipage}{0.50\linewidth}
			\leavevmode
			\includegraphics[width=1.\textwidth]{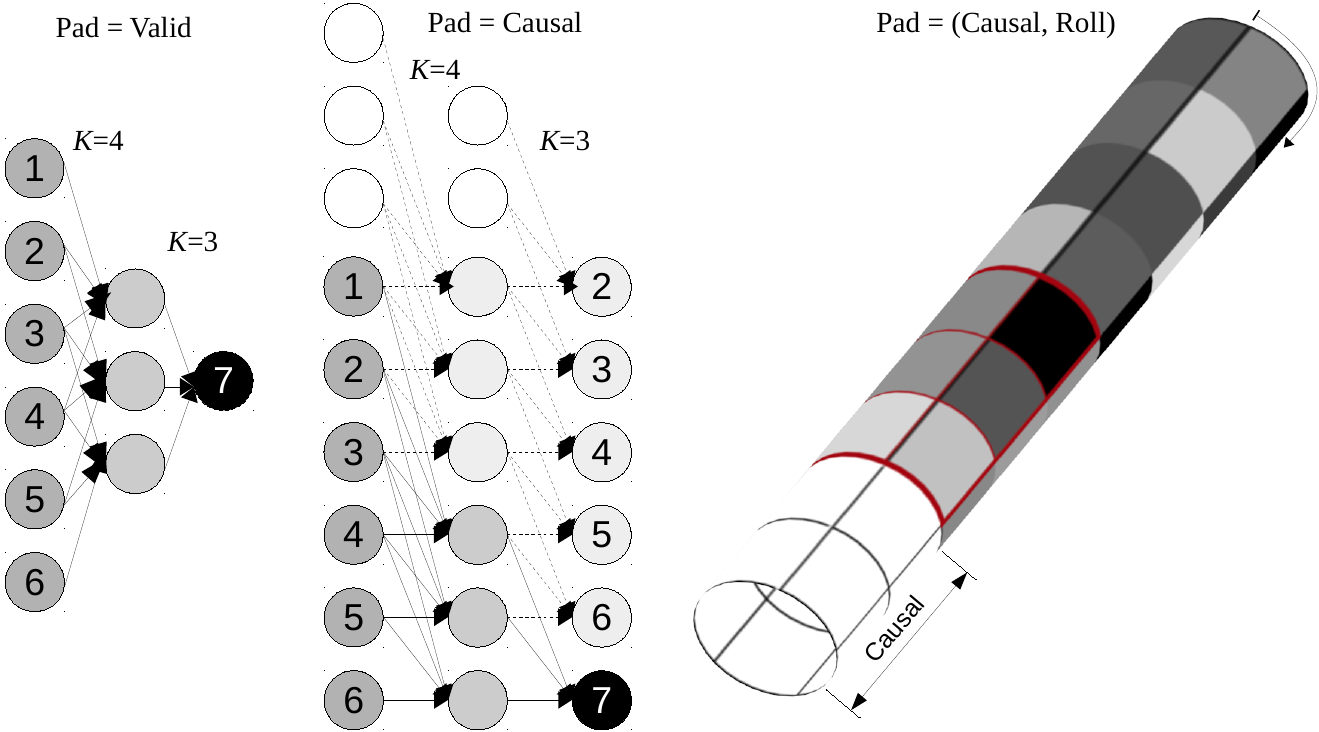}
			\hfill
		\end{minipage}
		\hspace{0.001\linewidth}
		\begin{minipage}{0.40\linewidth}
			\footnotesize
			\begin{tabular}{|c|c|c|c|}
				\hline
				& Roll & Variables & Roll \\
				\hline                                  
				\rotatebox[origin=l]{90}{Valid } &  &  &  \\
				\hline                                  
				&\fontfamily{qcr}\selectfont cdef&\fontfamily{qcr}\selectfont abcdef& \fontfamily{qcr}\selectfont abcd \\
				&\fontfamily{qcr}\selectfont ijkl&\fontfamily{qcr}\selectfont ghijkl&\fontfamily{qcr}\selectfont ghij\\
				\multirow{-3}{*}{\rotatebox[origin=l]{90}{Time }} &\fontfamily{qcr}\selectfont opqr&\fontfamily{qcr}\selectfont mnopqr&\fontfamily{qcr}\selectfont mnop\\
				\hline                                  
				& 			\fontfamily{qcr}\selectfont 0000&
				\fontfamily{qcr}\selectfont 000000&			\fontfamily{qcr}\selectfont 0000\\
				&\fontfamily{qcr}\selectfont 0000&
				\fontfamily{qcr}\selectfont 000000&			\fontfamily{qcr}\selectfont 0000\\
				\multirow{-3}{*}{\rotatebox[origin=l]{90}{Causal }}& & &\\
				\hline
			\end{tabular}
		\end{minipage}
	\end{center}
	\caption{On the left subplot, comparison behavior between valid and causal padding. On the right subplot, combination of causal and roll padding scheme for \gls{MTS} analysis with WaveNet 2D.}
	\label{fig-roll-causal-pad}
\end{figure}

A relevant \gls{DL} architecture that can be applied to \gls{MTS} problems is WaveNet, developed by Google DeepMind \citep{oord2016wavenet}. WaveNet was originally designed for audio signal generation. A key component in achieving this task was a sound classifier based on 1D convolutional layers. In the first two left subplots of Fig. \ref{fig-roll-causal-pad}, the output difference between using and not using causal padding in 1D convolutions can be observed. Causal padding ensures that the time steps of past information are preserved for the subsequent layer. In the context of \gls{MTS}, this requires that the number of zeros to be added before the beginning of all sequences is equal to $k-1$, where $k$ represents the size of the one-dimensional kernel for 1D convolutions. It is important to note that, for \gls{MTS} problems, each input variable is treated as a channel in 1D convolutions, and causal padding is uniformly applied across all channels.

WaveNet employs dilated convolutions to progressively expand the receptive field. In the time steps component, when using a dilation rate $dr$ (i.e., for $dr > 1$), the causal padding size is given by $dr \times (k-1)$. The residual block in the WaveNet architecture is executed multiple times according to a specified $depth$ in the network, with $N = \{1, \dots, depth\}$. Within this block, the dilation $dr$ of the \textit{sigmoid} and \textit{Tanh} convolutions applied to the time steps component increases exponentially according to the formula $dr = k^N$. As highlighted in Fig. \ref{fig-wavenet}, the third and final convolution within the residual block has $k=1$ and $dr=1$ to reduce dimensionality and manage model complexity. This layer is also referred to as the \textit{channel-wise pooling layer}. The standard WaveNet uses 1D convolutions, which can be adapted for \gls{MTS} problems by treating the input as a multichannel set of 1D sequences of variables. To explore the inclusion of roll padding in the variables component, the WaveNet architecture is extended to use 2D convolutions.

\subsubsection{WaveNet Extended with 2D Convolutions and Roll Padding}
\begin{figure}[htbp]
	\begin{center}
		\leavevmode
		\includegraphics[width=0.8\textwidth]{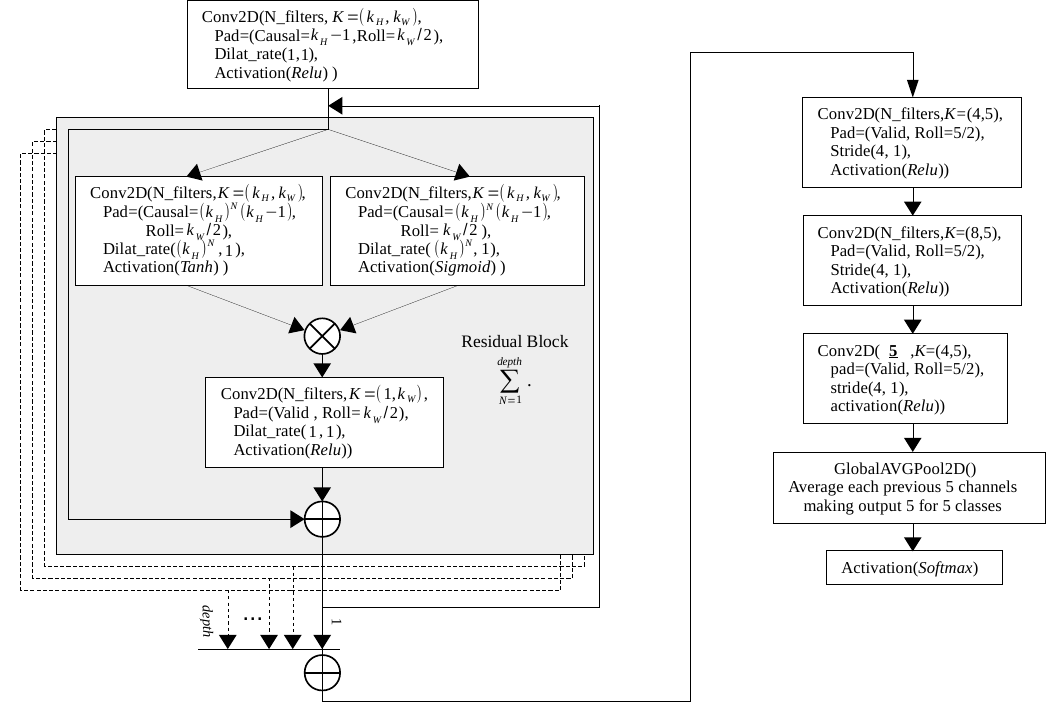} 
		\caption{WaveNet 2D architecture for \gls{MTS} classification using 2D convolutions with causal padding in the time steps component and roll padding in the variables component.}
		\label{fig-wavenet}
	\end{center}
\end{figure}

Fig. \ref{fig-wavenet} shows the extended WaveNet architecture, which operates with 2D maps instead of 1D multichannel inputs. The architecture preserves the standard WaveNet processing for the time steps component, using causal padding, while incorporating roll padding in the variables component. With the introduction of 2D convolutions, the kernel size is now defined as $(k_H, k_W)$. The $k_H$ component (i.e., the time steps component) is processed, with both $dr_H$ and causal padding applied to the time steps component, as described in Subsection \ref{sec-WaveNet-multi-head}. The combination of causal and roll padding is clarified in the last two right subplots of Fig. \ref{fig-roll-causal-pad}. In the second dimension (i.e., variables component), roll padding is applied, with a size determined by $k_W$. A roll padding size of ${k_W}/{2}$ is established, copying the opposite ${k_W}/{2}$ columns from the input map. For simplicity, odd fixed sizes in $k_W$ are assumed. No dilation rate is applied in the variables component (i.e., $dr_W=1$). In summary, the time steps component is processed in line with the basic WaveNet approach, while the variables component is processed using standard convolutional layers with roll padding. Finally, after adding skip connections, three 2D convolutional layers are included. In this scheme, a stride $\delta=(\delta_H, \delta_W)$ with $\delta_H > 1$ and $\delta_W = 1$ is used to downsample only the time steps dimension, rather than using pooling layers. The final convolution layer has $x$ filters ($x = 6$ in Fig. \ref{fig-wavenet}), generating $x$ feature maps, to which global average pooling is applied. This allows us to directly apply softmax to the $x$ resulting values for classification into $x$ classes.

\section{Results}\label{sec-4}

An end-to-end analysis was conducted for this case study, as a complete framework was established, covering everything from data gathering to market interaction. Table \ref{table-bf-final-stats} shows that the best result was achieved using the \gls{LSTM}-based model with multi-head attention, processing the attention weights of each variable time steps with Conv1D\footnote{Fig. \ref{fig-best-model-dia} in Appendix shows the \gls{DL} digram implementation for this best model.}. \gls{LSTM} models outperformed the other alternatives presented. This suggests that the inherent nature of the \gls{LSTM} architecture is well-suited to this type of data. Additionally, it is evident that the inclusion of roll padding improves the accuracy of both simple \gls{CNN} and WaveNet models compared to their base versions. However, due to the relatively small sample size in this case study, these \gls{CNN}-based models appear to be highly sensitive to overfitting. Further investigation of smaller \gls{CNN}-based solutions with better convergence is recommended.

\begin{table}[htbp]
	\begin{center}
		\small
		\begin{tabular}{|c||c|c|c|c|}	
			\hline
			Model & Min & Max& Mean & Variance   \\
			\hline
			\hline
			\gls{CNN} & 0.2342 & 0.2488& 0.2381& 3.6756e-05 \\
			\gls{CNN} ROLL & 0.2391& 0.2536& 0.2430&3.9674e-05\\
			\hline
			\textbf{\gls{LSTM}} & 0.2826 & 0.2874 & 0.2840&4.6675e-06  \\
			\textbf{\gls{LSTM} Std. Att}. &0.2681&0.2826&0.2720&3.9674e-05\\
			\textbf{\gls{LSTM} Att. Conv1D} & 0.2971&\textbf{0.3092}&0.3005&2.5088e-05 \\
			\hline
			\gls{ConvLSTM2D} 
			& 0.2584&0.2801&0.2671&7.1763e-05  \\
			\gls{ConvLSTM2D} Att. Conv2D 
			&0.2608&0.2705&0.2642&1.3419e-05    \\
			\hline
			WaveNet &0.2512&0.2681&0.2589&7.1180e-05  \\
			WaveNet2D ROLL & 0.2826&0.2922&0.2864&1.4518e-05  \\
			\hline
			
		\end{tabular}
		\caption{Accuracies obtained by 5 repetitive runs of the fitting process for each model applied to the case study.}
		\label{table-bf-final-stats}
	\end{center}
\end{table}

Table \ref{table-bf-best} presents the best \gls{DL} model, which achieves an accuracy of 30.92\%. Although this is only 11 percentage points above the baseline, it is important to consider that the goal of this problem is to generate a positive PL. Even with some incorrect predictions, a positive \gls{PL} can still be obtained. For instance, predicting a movement to the weak up class when the actual movement is a strong up class can still result in a positive \gls{PL}. 

\begin{table}[htbp]
	\begin{center}
		\begin{tabular}{|l|r||c|c|c|c|c||r|c|}
			\cline{3-7} 
			\multicolumn{2}{c||}{\multirow{2}{*}{}} & \multicolumn{5}{c||}{Predicted}& \multicolumn{2}{c}{}\\
			\cline{3-7} 
			\multicolumn{1}{c}{}&Classes  
			&{\rotatebox[origin=l]{90}{Strong Down}}& \rotatebox[origin=l]{90}{Weak Down} & \rotatebox[origin=l]{90}{Neutral} & \rotatebox[origin=l]{90}{Weak Up} & \rotatebox[origin=l]{90}{Strong Up} & \multicolumn{1}{c}{Recall(\%)} &\multicolumn{1}{c}{ACC (\%)} \\
			\hline
			\hline
			\multirow{6}{*}{\rotatebox[origin=l]{90}{Real}}
			& Strong Down & \cellcolor{green} 32 & \cellcolor{green}7  &  11 & \cellcolor{red} 9   & \cellcolor{red}10  & \cellcolor{green} 46.38 &  \multirow{6}{*}{\textbf{30.92}} \\ \cline{2-8} 
			& Weak Down & \cellcolor{green}39   &\cellcolor{green} 9 & 17   & \cellcolor{red}10   &\cellcolor{red} 11  & \cellcolor{green!30}10.47 &    \\ \cline{2-8} 
			& Neutral & 27   &  6  & 23 & 9   & 18  & 27.71 &   \\ \cline{2-8} 
			& Weak Up & \cellcolor{red}23   & \cellcolor{red} 13  &  19  &\cellcolor{green} 9 &\cellcolor{green} 27 & \cellcolor{green!30} 9.89 &     \\ \cline{2-8} 
			& Strong Up & \cellcolor{red}14   & \cellcolor{red} 8  &  15  & \cellcolor{green} 10  & \cellcolor{green}38  & \cellcolor{green} 44.71   &    \\ \hline \hline
			\multicolumn{1}{c}{}& \multicolumn{1}{c||}{Precision (\%)}    
			& \cellcolor{green}23.70   & \cellcolor{green!30} 20.93  &  27.06  & \cellcolor{green!30} 19.15  & \cellcolor{green}36.54  & 	\multicolumn{2}{c}{}      \\ 		\cline{3-7} 
		\end{tabular}
		\caption{Confusion matrix for \gls{LSTM}-based model with Conv1D multi-head attention on the validation dataset.}
		\label{table-bf-best}
	\end{center}
\end{table}

Table \ref{table-bf-best} confirms in green color all cases with a positive \gls{PL}. Similarly, cases that generate a negative \gls{PL} are indicated by the red color. The marginal convergence of the \gls{DL} model causes predicted values to approximate the main diagonal, drawing them towards the green cells and away from the red cells, which allows for a profitable model even with relatively low accuracy. For this specific case study, with 20\% of the data used for validation, the model predicts 173 trades with expected positive \gls{PL} and 98 trades with expected negative \gls{PL}, resulting in a total of 75 predicted trades with expected positive \gls{PL}. To fully assess its potential, the model needs to be evaluated in a production environment, as numerous factors can influence trade execution. This is done using a final test dataset of 30 days, which was not used during the modeling phase, to test the model's performance in a real-world scenario.

\begin{figure}[htbp]
	\begin{minipage}{0.497\linewidth}
		\leavevmode
		\includegraphics[width=1.\textwidth]{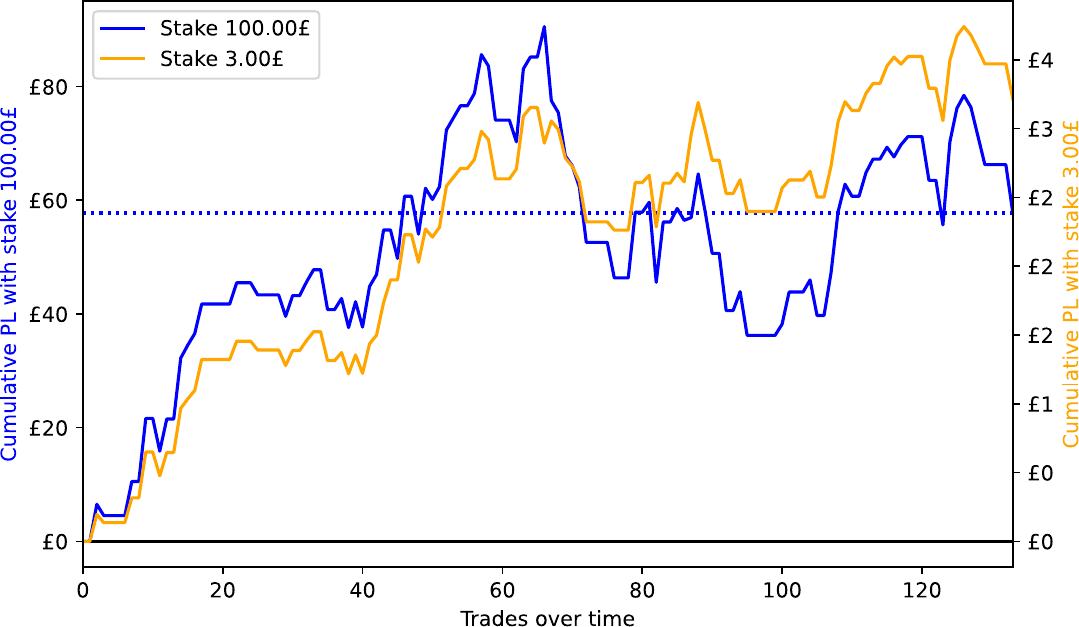}
		\hfill
	\end{minipage}
	\begin{minipage}{0.49\linewidth}
		\leavevmode
		\includegraphics[width=1.\textwidth]{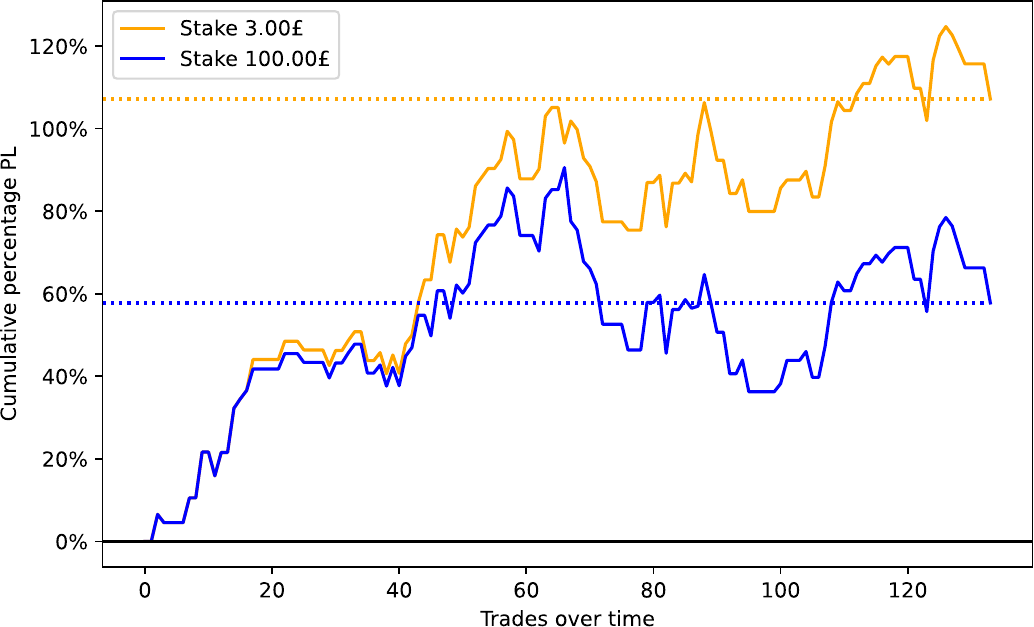}
		\hfill
		
	\end{minipage}
	\caption[Evolution of the PL during 30 days of trading using the best model, in the betting exchange case study.]{Evolution of the PL during 30 days of trading using the best model, of one category ($\#41$), with stakes of \pounds{3.00} and \pounds{100.00}. On the left subplot, absolute PL values. On the right subplot, relative PL values in relation to the investment stake.}
	\label{fig-results-pl-bf}
\end{figure}

Fig. \ref{fig-results-pl-bf} presents the cumulative \gls{PL} from the execution of trades in simulation for the final test dataset in one category. The left-hand subplot displays the absolute \gls{PL} achieved with stakes of \pounds{3.00} and \pounds{100.00}, while the right-hand subplot shows the relative \gls{PL}, calculated as a percentage of the initial investment, for the same stakes. By executing the same predicted trades, it is evident that using a stake of \pounds{3.00} generates a higher return on investment (ROI) compared to using \pounds{100.00}. This is attributed to the market's absorption capacity. Future studies should focus on evaluating the optimal stake policy to maximize the absolute \gls{PL}.

\begin{table}[!htbp]
	\begin{center}
		\small
		\begin{tabular}[t]{|c|c|c|}
			\hline 	Stake  &
			\pounds{3.00} & \pounds{100.00}\\ \hline \hline 
			Trades & 134  & 134 \\ \hline 
			Greens & 54 & 50\\  \hdashline
			\footnotesize Reaches target  &\footnotesize 14 &  13 \\ 
			\footnotesize Closes $]$null$;$target$[$ & \footnotesize 40 & 37\\ \hdashline
			Swings & 14 & 13  \\
			Trailing-Stops & 40 & 37 \\	\hline 
			Reds & 36 & 39 \\ \hdashline
			\footnotesize Reaches stop-loss & \footnotesize 13 & 14\\
			\footnotesize Breaks stop-loss & \footnotesize 3  & 5\\ 
			\footnotesize Closes $]$null$;$stop-loss$[$ & \footnotesize 23 &  25\\  \hdashline
			Swings & 10 &  11 \\
			Trailing-Stops & 26 & 28 \\	\hline 
			Null & 44  & 45\\ \hline \hline 
			Positive ticks & 134 & 129\\ \hdashline
			Swings & 31 &  28 \\
			Trailing-Stops & 103 & 101 \\  \hline 						
			Negative ticks &  87 & 100\\\hdashline
			Swings & 23 &  26 \\
			Trailing-Stops & 64 & 74 \\  \hline 										
		\end{tabular}
		\caption{Global trading simulation results with the model in production on the final test dataset.}
		\label{table-ticks}
	\end{center}
\end{table}

Table \ref{table-ticks} summarizes the number of executed trades, greens, reds, positive ticks, and negative ticks for the best \gls{DL} model. The number of trades refers to the total instances when a trading mechanism is instantiated. Greens represent the number of trades that close in profit, while reds indicate the number of trades that close in loss. The sum of greens and reds does not always equal the total number of trades, as there are cases where the trading mechanism closes a position at the same entry price, resulting in no profit or loss. This can occur when the trade reaches its timeout exposure and closes at the entry price. Additionally, if the opening bet is not matched during the opening time, the result is also null. Positive ticks represent the total ticks that result from profitable trades, while negative ticks account for the ticks resulting in losses. These are the key metrics used to evaluate the effectiveness of the trading policy.

Table \ref{table-bf-best-final-test-data} presents the confusion matrix for the model's predictions on the final test dataset, which generates the data in Table \ref{table-ticks}. The results are consistent; the number of trades instantiated corresponds to the number of times a class that results in an action is predicted, i.e., 134. The total expected number of green trades in Table \ref{table-bf-best-final-test-data} is 66, and reds are 41, which differs from the actual result in Table \ref{table-ticks}. This discrepancy can be attributed to instances where the model predicts a direction, but the actual class is Neutral, leading to trades with small positive or negative outcomes. Additionally, the classes are feature-engineered from the raw values representing the integral of price evolution during the prediction period. While this integral can indicate a consistent movement in one direction, rapid, aggressive price movements in the opposite direction can occur, contributing to the discrepancy between the expected and actual outcomes. As a plausible indicator of model resilience, the overall result ratios remain similar in both the production phase (Table \ref{table-bf-best-final-test-data}) and the validation dataset confusion matrices (Table \ref{table-bf-best}).

\begin{table}[htbp]
	\begin{center}
		\begin{tabular}{|l|r||c|c|c|c|c||r|c|}
			\cline{3-7} 
			\multicolumn{2}{c||}{\multirow{2}{*}{}} & \multicolumn{5}{c||}{Predicted}& \multicolumn{2}{c}{}\\
			\cline{3-7} 
			\multicolumn{1}{c}{}&Classes  
			&{\rotatebox[origin=l]{90}{Strong Down}}& \rotatebox[origin=l]{90}{Weak Down} & \rotatebox[origin=l]{90}{Neutral} & \rotatebox[origin=l]{90}{Weak Up} & \rotatebox[origin=l]{90}{Strong Up} & \multicolumn{1}{c}{Recall(\%)} &\multicolumn{1}{c}{ACC (\%)} \\
			\hline
			\hline
			\multirow{6}{*}{\rotatebox[origin=l]{90}{Real}}
			& Strong Down & \cellcolor{green} 16 & \cellcolor{green}3  &  2 & \cellcolor{red} 2   & \cellcolor{red}4  & \cellcolor{green} 59.26 &  \multirow{6}{*}{\textbf{28.26}} \\ \cline{2-8} 
			& Weak Down & \cellcolor{green}12   &\cellcolor{green} 6 & 1   & \cellcolor{red}5   &\cellcolor{red} 11  & \cellcolor{green!30}17.14 &    \\ \cline{2-8} 
			& Neutral & 14   &  4  &1 & 3   & 6  & 3.57 &   \\ \cline{2-8} 
			& Weak Up & \cellcolor{red}10   & \cellcolor{red} 2  &  0  &\cellcolor{green} 3 &\cellcolor{green} 8 & \cellcolor{green!30} 13 &     \\ \cline{2-8} 
			& Strong Up & \cellcolor{red}2   & \cellcolor{red} 5  &  0  & \cellcolor{green} 5  & \cellcolor{green}13  & \cellcolor{green} 52   &    \\ \hline \hline
			\multicolumn{1}{c}{}& \multicolumn{1}{c||}{Precision (\%)}    
			& \cellcolor{green}29.63   & \cellcolor{green!30} 30  &  25  & \cellcolor{green!30} 16.67  & \cellcolor{green}30.95  & 	\multicolumn{2}{c}{}      \\ 		\cline{3-7} 
		\end{tabular}
		\caption{Confusion matrix for the LSTM-based model with Conv1D multi-head attention, i.e., best model, on the final test dataset.}
		\label{table-bf-best-final-test-data}
	\end{center}
\end{table}

\section{Managerial implications}\label{sec-5}

Results of this study demonstrate the feasibility and potential of leveraging \gls{DL} techniques, particularly \gls{LSTM}-based models with multi-head attention and Conv1D processing, for forecasting short-term price movements in the Betfair UK to Win Horse Racing market. The systematic approach to feature engineering, model training, and production testing provides valuable insights into predictive performance and practical implementation of such models in real-world trading environments. Therefore, managers should consider implementing \gls{LSTM}-based models with multi-head attention and Conv1D processing, for forecasting price movements, as these models are effective at capturing sequential dependencies in market data, crucial for accurately predicting short-term price fluctuations.

The study confirms that \gls{LSTM}-based architectures outperform alternative models, indicating their structure is well-suited for capturing sequential dependencies in market depth data. The inclusion of multi-head convolutional attention mechanisms further enhances the model's ability to focus on key patterns in time-series data, improving accuracy. Hence, managers should prioritize integrating this \gls{DL} technique over simpler ones, as they have been proven to deliver superior predictive performance, giving a competitive edge in dynamic markets. Additionally, multi-head convolutional attention mechanisms allow to refine trading algorithms, enabling them to better identify critical patterns in market data, which improves prediction accuracy. Additionally, the introduction of roll padding in \gls{CNN}-based models contributes to addtional performance gains. This implies that, from a managerial perspective, investing in continuous model refinement and newer techniques will ensure predictive performance remains robust, especially when adapting to changing market conditions.

Beyond the classification accuracy of 30.92\%, which is significantly higher than the 20\% expected from random choices, this research emphasizes the model's ability to generate a positive \gls{PL}. Results show that the model consistently predicts more trades with expected positive \gls{PL} than those with expected negative \gls{PL}, reinforcing its potential for profitable trading. Consequently, it is important to focus not just on accuracy but also on profitability metrics, ensuring that the model's predictions result in real-world gains. Aligning trading strategies with profitability measures and risk-adjusted returns will help in translating predictive success into tangible financial outcomes.

Furthermore, analysis of trade outcomes over a 30-day final test dataset reveals the model maintains a favorable expected \gls{PL}, though further refinements could improve its performance. Results also highlight the importance of market absorption capacity in determining optimal stake size. Managers should consider adaptive stake sizing strategies based on liquidity constraints, as smaller stakes of \pounds3.00 yield higher returns compared to larger stakes of \pounds100.00. This outcome confirms the influential role of liquidity in exchange markets, which impacts trade execution and profitability. Developing flexible stake-sizing strategies that adapt to market conditions will ensure that trading is optimized without overexposing the system to risks caused by liquidity issues.

Another key observation is the discrepancy between expected and actual numbers of positive and negative trades. This is mainly due to instances where the model predicts a direction but encounters volatile market behavior that diverges from the expected trend. These inconsistencies reinforce the importance of real-time market dynamics, order execution efficiency, and slippage in trading performance. As such, managers should ensure that real-time market conditions, such as volatility and execution efficiency, are continuously monitored to reduce slippage and improve trade execution. Improving order routing systems and adjusting trades dynamically could minimize discrepancies between predicted and actual outcomes.

\section{Conclusions}\label{sec-6}

This study advances automated trading strategies by integrating innovative convolutional attention mechanisms and a specialized padding method for time series forecasting. Specifically, it introduces novel enhancements to \gls{DL} \gls{NN} architectures, including roll padding, multi-head attention with Conv1D for \gls{LSTM}-based models, and multi-head attention with Conv2D and roll padding for \gls{ConvLSTM2D}-based models. Unlike previous studies, the practical implementation extends beyond these technical developments. By focusing on the UK to Win Horse Racing market during the pre-live stage of the world's leading betting exchange, this research proposes a comprehensive end-to-end framework for predicting price movements while emphasizing the importance of model robustness in real-world trading environments. These enhancements significantly improve the learning process, demonstrating the potential of \gls{DL} approaches when combined with domain expertise in betting exchange markets. At the same time, the study acknowledges challenges such as market volatility, data limitations, and model interpretability. Ultimately, this research opens new avenues for future studies in automated trading systems and contributes to the broader field of financial and sports market analytics.

This study presents several important findings. Firstly, \gls{LSTM}-based models with multi-head attention are more effective than alternative architectures when it comes to predicting short-term price movements. Secondly, while roll padding enhances \gls{CNN}-based models, it does not completely address the overfitting challenges arising from limited training data. Thirdly, focusing solely on classification accuracy is not an adequate measure of trading performance; instead, the ability to generate a net positive \gls{PL} is a more meaningful benchmark for success. Lastly, smaller stake sizes tend to yield higher returns, emphasizing the importance of liquidity and market absorption in optimizing trading strategies. 

As such, in terms of scholarly implications, this study introduces a data-driven framework for predicting price changes in betting exchange markets, reducing dependence on expert-driven approaches. From a managerial perspective, findings suggest that managers should prioritize refining and leveraging \gls{DL} techniques, particularly \gls{LSTM} models with multi-head convolutional attention mechanisms and roll padding, while optimizing trading strategies for profitability, liquidity constraints, and execution efficiency. Continuous model evaluation and adaptability to market dynamics remain critical for sustaining competitive advantages in real-time trading environments.

Despite the effort to make a valuable contribution, the study is not without its limitations. Future research could refine model convergence, optimize stake policies, and improve adaptability to real-world market fluctuations. Further work may involve expanding the dataset to enhance model generalization, developing adaptive trading strategies based on real-time market conditions, and exploring alternative techniques to dynamically optimize trade execution. Additionally, integrating other risk management strategies to mitigate potential losses will be crucial for the practical deployment of such systems in high-frequency trading environments.

\small{
	\bibliography{PAPER-FINAL}
}

\section*{Appendix}

\setcounter{table}{0}
\setcounter{figure}{0}
\setcounter{footnote}{0}
\setcounter{equation}{0}

\renewcommand{\thetable}{A\arabic{table}}
\renewcommand{\thefigure}{A\arabic{figure}}
\renewcommand{\thefootnote}{A\arabic{footnote}}
\renewcommand{\theequation}{A\arabic{equation}}

\begin{figure}[H]
	\centering
	\begin{subfigure}{0.90\textwidth}
		\centering
		\includegraphics[width=1\textwidth]{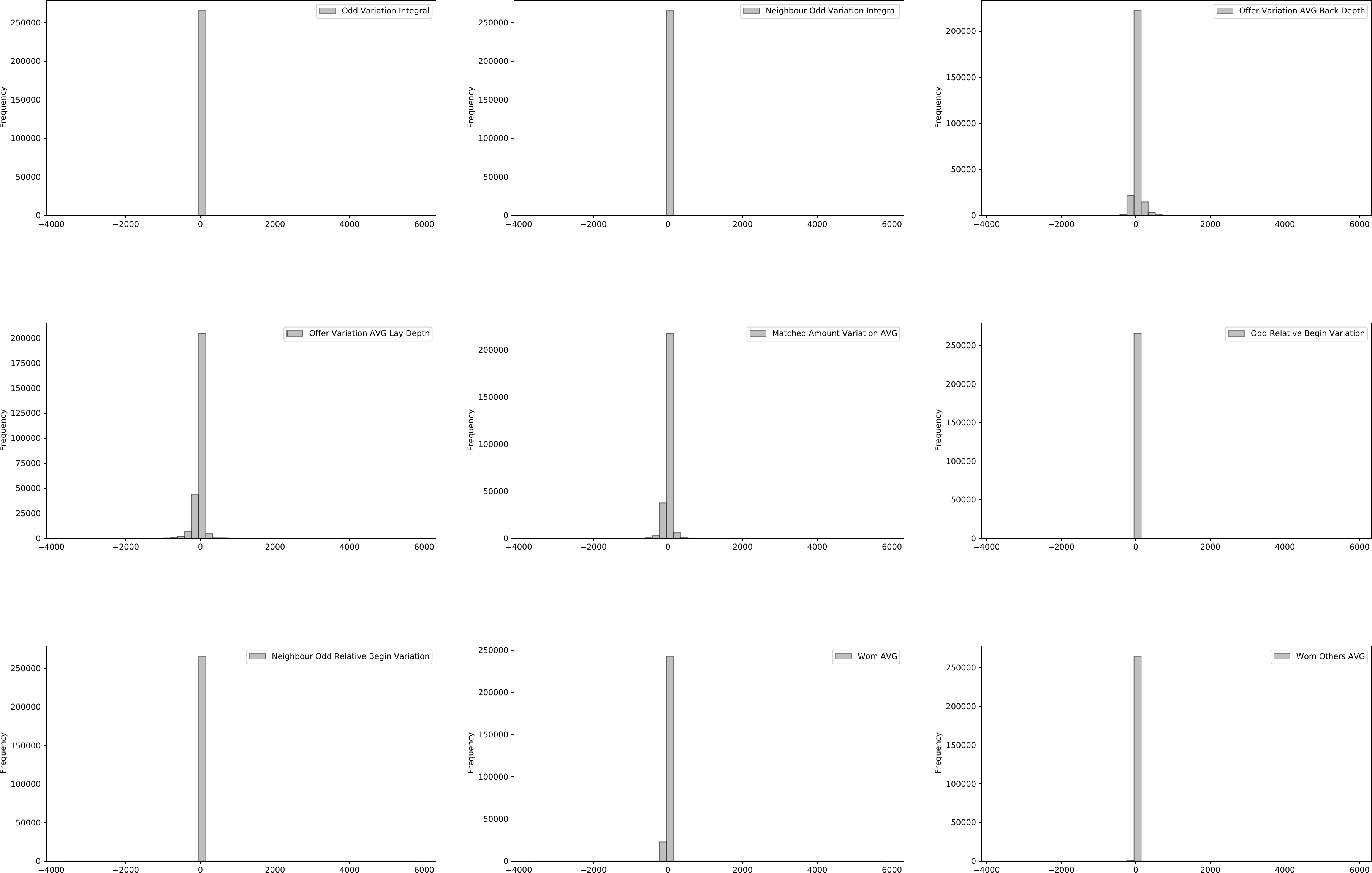}
		\label{fig:up-histo-horses}
	\end{subfigure}
	\noindent\rule{\textwidth}{1pt}
	\hfill\break
	\begin{subfigure}{0.90\textwidth}
		\centering
		\includegraphics[width =1 \textwidth]{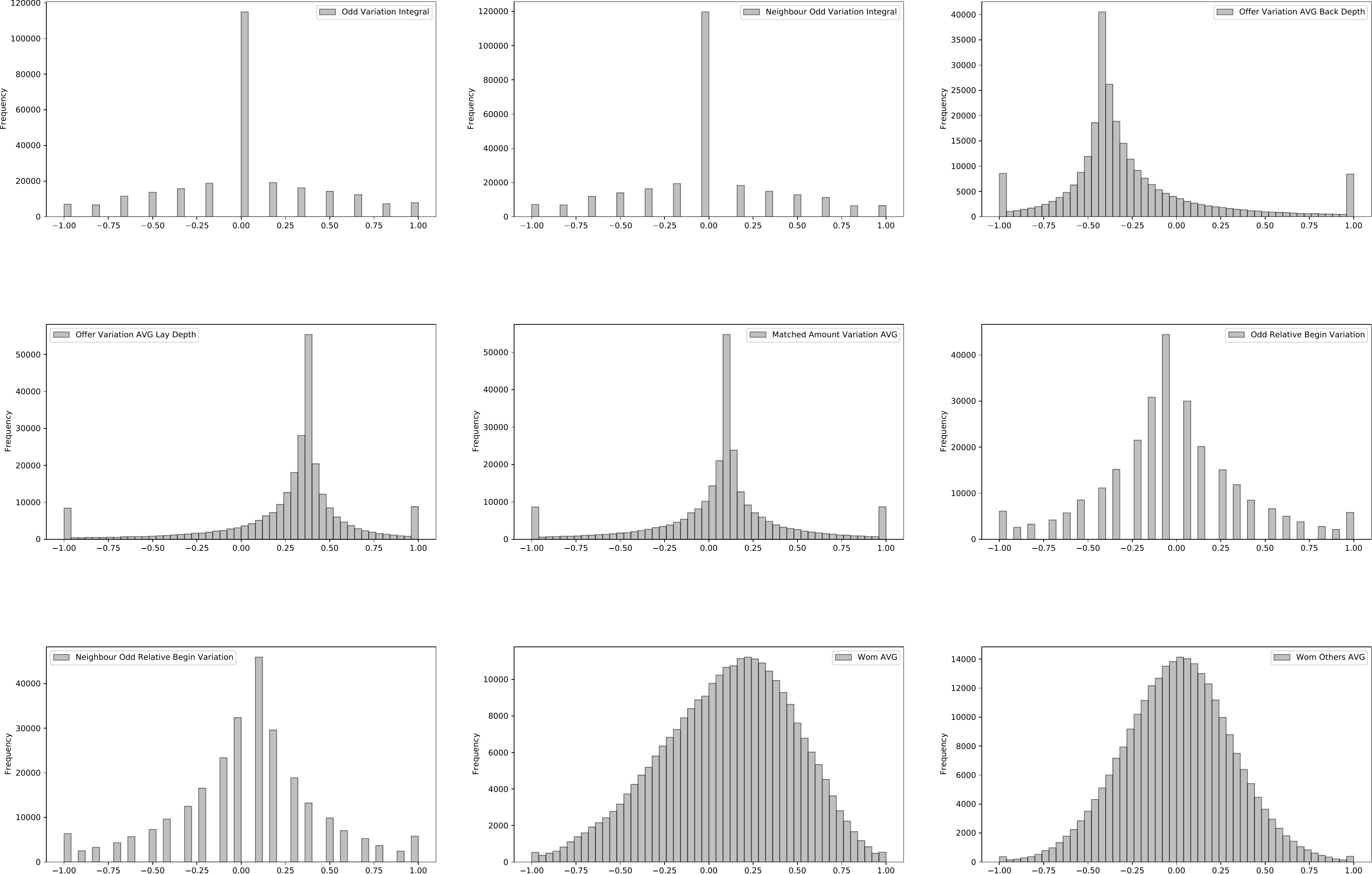}
		\label{fig:down-histo-horses}
	\end{subfigure}
	\caption[Example of histogram re-scaling for the normalization process on all indicator in the betting exchange case study.]{Example of histogram re-scaling with truncated tails at $10\%$ level to find min-max values for input normalization. The illustration represents inputs on category $\#41$ of the rule-based system index of the betting exchange horse race markets case study. Top 9 subplots are raw data and bottom 9 subplots are the result of applying this technique.}
	\label{fig-hist-horses-apx}
\end{figure}

\begin{sidewaysfigure}[htbp] 
	\begin{center}
				\leavevmode
		\includegraphics[width=1\textwidth]{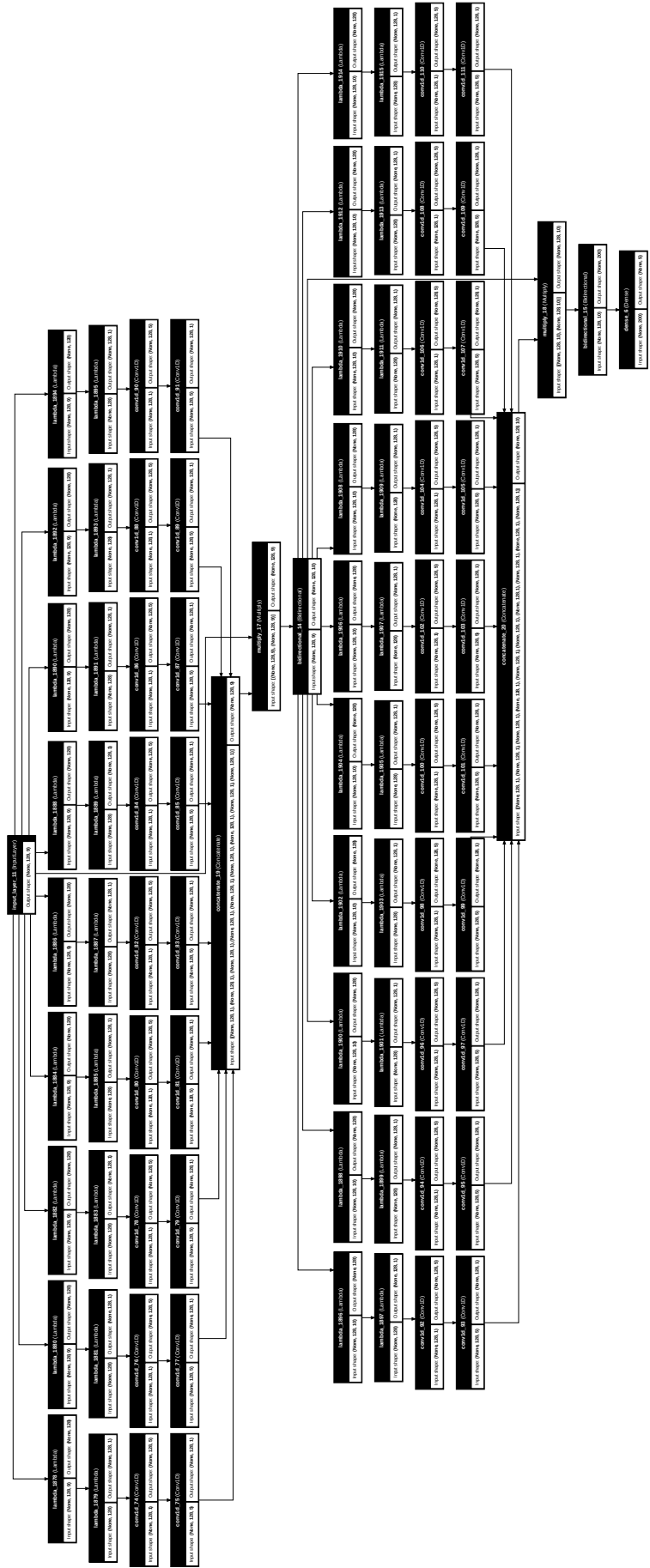} 
		\caption{Best model diagram - Multihead Convolutional Attention 1D with bidirectional \gls{LSTM}. We plot the model with only 2 blocks in depth for simplicity; the model used had 3 blocks with 3 stacked bidirectional \glspl{LSTM}. Also, in this diagram, the first bidirectional \gls{LSTM} only uses 5 recurrent cells (returning 10 sequences due to bidirection) for simplicity. For the case study, 50, 20, 5 recurrent cells were used in depth, respectively.}
		\label{fig-best-model-dia}

	\end{center}
\end{sidewaysfigure}

\begin{sidewaystable}[htbp] 
	\begin{center}
		
		\fontsize{5}{7.7}\selectfont 
		\begin{tabular}{ | l | l | l | l | l | l | l | l | l | l | l | l | l | l | l | l | l | l | l | }
			\hline
			\textbf{P\&L} & \textbf{TM} & \textbf{END\_STATE} & \textbf{EVENT} & \textbf{RUNNER} & \textbf{VOLUME} & \textbf{N\_R} & \textbf{ENTR} & \textbf{TARG} & \textbf{STO} & \textbf{DIR} & \textbf{T\_P} & \textbf{T\_L} & \textbf{PT\_P} & \textbf{PT\_L} & \textbf{O\_ODD} & \textbf{C\_ODD} & \textbf{O\_AM} & \textbf{CAT\_AM} \\ \hline
			\pounds0,00 & 2 & NOT\_OPEN & Ham\_2nd\_Sep & Nightster & 38189.27 & 8 & 5.5 & 6.2 & 5.1 & LB & 6 & 4 & \pounds11,29 & -\pounds7,84  & 0 & 0 & \pounds0,00 & \pounds0,00 \\ \hline
			\pounds0,00 & 2 & NOT\_OPEN & Ham\_2nd\_Sep & Tectonic & 18415.76 & 8 & 4.5 & 3.95 & 4.9 & BL & 6 & 4 & \pounds13,92 & -\pounds8,16  & 0 & 0 & \pounds0,00 & \pounds0,00 \\ \hline
			\pounds6,52 & 1 & CLOSED & FfosL\_2nd\_Sep & Mccool\_Bannanas & 23465.96 & 6 & 4.9 & 4.6 & 5.2 & BL & 3 & 3 & \pounds6,52 & -\pounds5,77 & 4.9 & 4.6 & \pounds100,00 & \pounds106,52 \\ \hline
			-\pounds2,00 & 1 & CLOSED & Ham\_2nd\_Sep & King\_Of\_Paradise & 15501.31 & 6 & 5.1 & 5.5 & 4.7 & LB & 4 & 4 & \pounds7,27 & -\pounds8,51 & 5.1 & 5 & \pounds100,00 & \pounds102,00 \\ \hline
			\pounds0,00 & 1 & NOT\_OPEN & FfosL\_2nd\_Sep & Cabuchon & 22716.31 & 7 & 4.4 & 4.1 & 4.7 & BL & 3 & 3 & \pounds7,32 & -\pounds6,38  & 0 & 0 & \pounds0,00 & \pounds0,00 \\ \hline
			\pounds0,00 & 1 & CLOSED & FfosL\_2nd\_Sep & Men\_Dont\_Cry & 24814.14 & 7 & 4.5 & 4.9 & 4.1 & LB & 4 & 4 & \pounds8,16 & -\pounds9,76  & 4.5 & 4.5 & \pounds100,00 & \pounds100,00 \\ \hline
			\pounds0,00 & 2 & NOT\_OPEN & Ham\_2nd\_Sep & George\_Fenton & 20976.69 & 9 & 4.9 & 4.3 & 5.3 & BL & 6 & 4 & \pounds13,95 & -\pounds7,55  & 0 & 0 & \pounds0,00 & \pounds0,00 \\ \hline
			\pounds6,00 & 1 & CLOSED & Brig\_2nd\_Sep & Admiralofthesea & 20249.86 & 9 & 5.3 & 5 & 5.6 & BL & 3 & 3 & \pounds6,00 & -\pounds5,36  & 5.3 & 5 & \pounds100,00 & \pounds106,00 \\ \hline
			\pounds0,00 & 1 & CLOSED & Muss\_3rd\_Sep & Mishaal & 21280.96 & 9 & 5 & 4.7 & 5.3 & BL & 3 & 3 & \pounds6,38 & -\pounds5,66  & 5 & 5 & \pounds100,00 & \pounds100,00 \\ \hline
			\pounds11,11 & 2 & CLOSED & Good\_3rd\_Sep & Deeds\_Not\_Words & 21824.36 & 6 & 6 & 5.4 & 6.8 & BL & 6 & 4 & \pounds11,11 & -\pounds11,76 & 6 & 5.4 & \pounds100,00 & \pounds111,11 \\ \hline
			\pounds0,00 & 1 & CLOSED & Good\_3rd\_Sep & Argent\_Knight & 45269.49 & 9 & 5.6 & 5.3 & 5.9 & BL & 3 & 3 & \pounds5,66 & -\pounds5,08  & 5.6 & 5.6 & \pounds100,00 & \pounds100,00 \\ \hline
			-\pounds5,77 & 2 & CLOSED & Good\_3rd\_Sep & Minority\_Interest & 15054.53 & 10 & 4.9 & 4.3 & 5.3 & BL & 6 & 4 & \pounds13,95 & -\pounds7,55  & 4.9 & 5.2 & \pounds100,00 & \pounds94,23 \\ \hline
			\pounds5,66 & 2 & CLOSED & Good\_3rd\_Sep & Swift\_Blade & 19614.89 & 10 & 5.6 & 5 & 6 & BL & 6 & 4 & \pounds12,00 & -\pounds6,67  & 5.6 & 5.3 & \pounds100,00 & \pounds105,66 \\ \hline
			\pounds0,00 & 2 & NOT\_OPEN & Ling\_3rd\_Sep & Prospera & 30635.19 & 7 & 4.8 & 5.4 & 4.4 & LB & 6 & 4 & \pounds11,11 & -\pounds9,09  & 0 & 0 & \pounds0,00 & \pounds0,00 \\ \hline
			\pounds10,71 & 2 & CLOSED & Ling\_3rd\_Sep & Rock\_God & 29429.46 & 7 & 5 & 5.6 & 4.6 & LB & 6 & 4 & \pounds10,71 & -\pounds8,70  & 5 & 5.6 & \pounds100,00 & \pounds89,29 \\ \hline
			\pounds2,27 & 2 & CLOSED & Bath\_4th\_Sep & Kakapuka & 31543.94 & 7 & 4.5 & 3.95 & 4.9 & BL & 6 & 4 & \pounds13,92 & -\pounds8,16  & 4.5 & 4.4 & \pounds100,00 & \pounds102,27 \\ \hline
			\pounds2,04 & 2 & CLOSED & Bath\_4th\_Sep & Dreams\_Of\_Glory & 31036.05 & 7 & 4.8 & 5.4 & 4.4 & LB & 6 & 4 & \pounds11,11 & -\pounds9,09  & 4.8 & 4.9 & \pounds100,00 & \pounds97,96 \\ \hline
			\pounds5,20 & 2 & CLOSED & Bath\_4th\_Sep & Devon\_Diva & 6343.06 & 6 & 4.3 & 3.85 & 4.7 & BL & 6 & 4 & \pounds11,69 & -\pounds8,51  & 4.3 & 4 & \pounds69,27 & \pounds74,47 \\ \hline
			\pounds0,00 & 2 & CLOSED & Kemp\_4th\_Sep & For\_Posterity & 17928.36 & 8 & 4.6 & 4 & 5 & BL & 6 & 4 & \pounds15,00 & -\pounds8,00  & 4.6 & 4.6 & \pounds100,00 & \pounds100,00 \\ \hline
			\pounds0,00 & 2 & NOT\_OPEN & Salis\_5th\_Sep & Mysterious\_Man & 29961.6 & 6 & 4.2 & 3.8 & 4.6 & BL & 6 & 4 & \pounds10,53 & -\pounds8,70 & 0 & 0 & \pounds0,00 & \pounds0,00 \\ \hline
			\pounds0,00 & 2 & NOT\_OPEN & Salis\_5th\_Sep & New\_Rich & 14194.09 & 6 & 4.2 & 3.8 & 4.6 & BL & 6 & 4 & \pounds10,53 & -\pounds8,70  & 0 & 0 & \pounds0,00 & \pounds0,00 \\ \hline
			\pounds0,00 & 1 & NOT\_OPEN & Salis\_5th\_Sep & Catchanova & 19254.96 & 7 & 4.2 & 3.95 & 4.5 & BL & 3 & 3 & \pounds6,33 & -\pounds6,67  & 0 & 0 & \pounds0,00 & \pounds0,00 \\ \hline
			\pounds3,74 & 1 & CLOSED & Salis\_5th\_Sep & South\_Cape & 17128.67 & 7 & 4.3 & 4.7 & 3.95 & LB & 4 & 4 & \pounds8,51 & -\pounds8,86  & 4.3 & 4.47 & \pounds100,00 & \pounds96,26 \\ \hline
			-\pounds0,00 & 1 & CLOSED & Newc\_6th\_Sep & Red\_Pike & 17883.92 & 10 & 4.3 & 4.7 & 3.95 & LB & 4 & 4 & \pounds8,51 & -\pounds8,86  & 4.3 & 4.3 & \pounds100,00 & \pounds100,00 \\ \hline
			\pounds0,00 & 2 & CLOSED & Newc\_6th\_Sep & Noble\_Asset & 33583.39 & 9 & 4.3 & 4.9 & 3.95 & LB & 6 & 4 & \pounds12,24 & -\pounds8,86 & 4.3 & 4.3 & \pounds100,00 & \pounds100,00 \\ \hline
			-\pounds2,13 & 2 & CLOSED & Newc\_6th\_Sep & Pure\_Impressions & 11972.8 & 9 & 4.6 & 4 & 5 & BL & 6 & 4 & \pounds15,00 & -\pounds8,00 & 4.6 & 4.7 & \pounds100,00 & \pounds97,87 \\ \hline
			\pounds0,00 & 1 & CLOSED & Hayd\_6th\_Sep & Ashpan\_Sam & 20740.66 & 8 & 4.3 & 4.7 & 3.95 & LB & 4 & 4 & \pounds8,51 & -\pounds8,86  & 4.3 & 4.3 & \pounds93,25 & \pounds93,25 \\ \hline
			\pounds0,00 & 1 & NOT\_OPEN & Hayd\_6th\_Sep & Anomaly & 19408.38 & 8 & 5.8 & 5.5 & 6.2 & BL & 3 & 3 & \pounds5,45 & -\pounds6,45  & 0 & 0 & \pounds0,00 & \pounds0,00 \\ \hline
			\pounds0,00 & 2 & NOT\_OPEN & Kemp\_7th\_Sep & Masterstroke & 26429.99 & 10 & 6 & 5.4 & 6.8 & BL & 6 & 4 & \pounds11,11 & -\pounds11,76 & 0 & 0 & \pounds0,00 & \pounds0,00 \\ \hline
			-\pounds3,77 & 2 & CLOSED & Ascot\_7th\_Sep & Steventon\_Star & 10016.69 & 7 & 5.1 & 4.5 & 5.5 & BL & 6 & 4 & \pounds13,33 & -\pounds7,27  & 5.1 & 5.3 & \pounds100,00 & \pounds96,23 \\ \hline
			\pounds3,64 & 2 & CLOSED & Ascot\_7th\_Sep & Forgive & 19005.23 & 8 & 5.3 & 5.9 & 4.9 & LB & 6 & 4 & \pounds10,17 & -\pounds8,16  & 5.3 & 5.5 & \pounds100,00 & \pounds96,36 \\ \hline
			\pounds0,00 & 1 & NOT\_OPEN & Wolv\_7th\_Sep & Lord\_Buffhead & 48688.37 & 10 & 5.8 & 6.4 & 5.4 & LB & 4 & 4 & \pounds9,38 & -\pounds7,41 & 0 & 0 & \pounds0,00 & \pounds0,00 \\ \hline
			\pounds2,44 & 2 & CLOSED & Font\_8th\_Sep & Brough\_Academy & 13833.73 & 7 & 4.2 & 3.8 & 4.6 & BL & 6 & 4 & \pounds10,53 & -\pounds8,70  & 4.2 & 4.1 & \pounds100,00 & \pounds102,44 \\ \hline
			\pounds2,13 & 2 & CLOSED & Font\_8th\_Sep & Chilworth\_Screamer & 18831.11 & 7 & 4.8 & 4.2 & 5.2 & BL & 6 & 4 & \pounds14,29 & -\pounds7,69 & 4.8 & 4.7 & \pounds100,00 & \pounds102,13 \\ \hline
			-\pounds0,00 & 2 & CLOSED & Font\_8th\_Sep & The\_Tracey\_Shuffle & 44132.1 & 7 & 4.1 & 4.7 & 3.85 & LB & 6 & 4 & \pounds12,77 & -\pounds6,49  & 4.1 & 4.1 & \pounds100,00 & \pounds100,00 \\ \hline
			-\pounds7,02 & 1 & CLOSED & Font\_8th\_Sep & Ovilia & 47113.2 & 9 & 5.3 & 5 & 5.6 & BL & 3 & 3 & \pounds6,00 & -\pounds5,36  & 5.3 & 5.7 & \pounds100,00 & \pounds92,98 \\ \hline
			\pounds0,00 & 2 & NOT\_OPEN & Hunt\_9th\_Sep & Brimham\_Boy & 19783.33 & 9 & 6 & 7.2 & 5.6 & LB & 6 & 4 & \pounds16,67 & -\pounds7,14  & 0 & 0 & \pounds0,00 & \pounds0,00 \\ \hline
			\pounds1,96 & 1 & CLOSED & Hunt\_9th\_Sep & Tiny\_Tenor & 10597.64 & 6 & 5 & 5.4 & 4.6 & LB & 4 & 4 & \pounds7,41 & -\pounds8,70  & 5 & 5.1 & \pounds100,00 & \pounds98,04 \\ \hline
			-\pounds5,13 & 2 & CLOSED & Perth\_9th\_Sep & Rathmoyle\_House & 15105.65 & 6 & 4.1 & 4.7 & 3.85 & LB & 6 & 4 & \pounds12,77 & -\pounds6,49  & 4.1 & 3.9 & \pounds100,00 & \pounds105,13 \\ \hline
			\pounds4,55 & 2 & CLOSED & Hunt\_9th\_Sep & Getaway\_Car & 43158.07 & 7 & 4.6 & 4 & 5 & BL & 6 & 4 & \pounds15,00 & -\pounds8,00  & 4.6 & 4.4 & \pounds100,00 & \pounds104,55 \\ \hline
			-\pounds4,44 & 2 & CLOSED & Brig\_9th\_Sep & Arlecchino & 34162.89 & 6 & 4.3 & 3.85 & 4.7 & BL & 6 & 4 & \pounds11,69 & -\pounds8,51  & 4.3 & 4.5 & \pounds100,00 & \pounds95,56 \\ \hline
			\pounds7,14 & 1 & CLOSED & Brig\_9th\_Sep & Kamchatka & 8723.94 & 6 & 4.5 & 4.2 & 4.8 & BL & 3 & 3 & \pounds7,14 &-\pounds6,25  & 4.5 & 4.2 & \pounds100,00 & \pounds107,14 \\ \hline
			\pounds2,08 & 2 & CLOSED & Leic\_10th\_Sep & Tatlisu & 32783.13 & 7 & 4.9 & 4.3 & 5.3 & BL & 6 & 4 & \pounds13,95 & -\pounds7,55  & 4.9 & 4.8 & \pounds100,00 & \pounds102,08 \\ \hline
			\pounds7,84 & 2 & CLOSED & Worc\_10th\_Sep & Gud\_Day & 16815.93 & 6 & 4.7 & 5.3 & 4.3 & LB & 6 & 4 & \pounds11,32 & -\pounds9,30  & 4.7 & 5.1 & \pounds100,00 & \pounds92,16 \\ \hline
			\pounds0,00 & 2 & NOT\_OPEN & Bev\_10th\_Sep & Hadaj & 19770.65 & 8 & 4.3 & 3.85 & 4.7 & BL & 6 & 4 & \pounds11,69 & -\pounds8,51  & 0 & 0 & \pounds0,00 & \pounds0,00 \\ \hline
			-\pounds5,00 & 1 & CLOSED & Bev\_10th\_Sep & Bondi\_Beach\_Boy & 16922.4 & 8 & 5.7 & 5.4 & 6 & BL & 3 & 3 & \pounds5,56 & -\pounds5,00  & 5.7 & 6 & \pounds100,00 & \pounds95,00 \\ \hline
			\pounds10,94 & 2 & CLOSED & Bev\_10th\_Sep & Dubai\_Dynamo & 20403.49 & 8 & 5.7 & 6.6 & 5.3 & LB & 6 & 4 & \pounds13,64 & -\pounds7,55  & 5.7 & 6.4 & \pounds100,00 & \pounds89,06 \\ \hline
			\pounds0,00 & 2 & NOT\_OPEN & Bev\_10th\_Sep & Starlit\_Cantata & 35090.79 & 9 & 5.1 & 4.5 & 5.5 & BL & 6 & 4 & \pounds13,33 & -\pounds7,27  & 0 & 0 & \pounds0,00 & \pounds0,00 \\ \hline
			-\pounds6,67 & 2 & CLOSED & Uttox\_11th\_Sep & Teak & 19655.05 & 7 & 5.6 & 5 & 6 & BL & 6 & 4 & \pounds12,00 & -\pounds6,67  & 5.6 & 6 & \pounds100,00 & \pounds93,33 \\ \hline

			... & ... & ... & ...  & ... & ... & ... & ... & ... & ... & ... & ... & .. & ... & ... & .. & ... & ... & ... \\ \hline
		\end{tabular}

		\caption{Logs segment showing the output fields of trading mechanisms (TM 2 - Trailing-stop and 1 - Swing) and parametrization according to the DL models prediction, during several races. The base stake used is  \pounds100.00.}
		\label{table-horse-final-output}
	\end{center}
\end{sidewaystable}	

\end{document}